%% file: main.tex
\documentclass[a4paper,11pt]{article}
\pdfoutput=1

\usepackage{jheppub}
\usepackage[T1]{fontenc}
\usepackage{amsfonts,bm}
\usepackage{booktabs}
\usepackage{xcolor}
\usepackage{braket}
\usepackage{mathtools}
\usepackage{multirow}
\usepackage{placeins}

\input{diagram_setup.tex}
\input{metadata.tex}
\input{macros.tex}
\input{numbers.tex}

\hypersetup{pdftitle={\PaperTitlePDF},pdfauthor={\PaperAuthor}}

\title{\boldmath Two-loop $gg\to AA$ amplitudes near threshold and the double parton scattering singularity}

\author{Arunima Bhattacharya\,\href{https://orcid.org/0000-0003-4409-0901}{\includegraphics[scale=0.05]{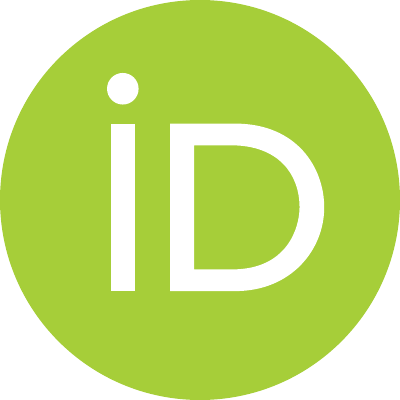}}}
\affiliation{Theory Division, IFIC, University of Valencia--CSIC,\\
E-46980 Paterna, Valencia, Spain}
\emailAdd{\PaperEmail}

\abstract{
The two-loop amplitudes for pseudoscalar pair production in gluon fusion, $gg\to AA$, with each pseudoscalar emitted from its own effective gluon vertex, are known in closed form, but large cancellations near threshold obscure their analytic structure. We expand the IR-finite amplitudes in the pseudoscalar velocity $\beta_A$ through $\beta_A^6$ at fixed scattering angle. All inverse powers of $\beta_A$ and the odd powers through $\beta_A^5$ cancel, and at leading power the angular dependence enters only through $\ln(2p_T/m_A)$. We identify this logarithm with the double parton scattering (DPS) singularity: with the overall constant fixed at threshold, the factorisation formula of Gaunt and Stirling reproduces the velocity dependence of its one-loop coefficient in the forward limit exactly, and a one-loop $gg\to HH$ analysis confirms the corresponding CP-even helicity sum. At two loops the equal-helicity amplitude grows as $\ln^3(2p_T/m_A)$, and the ratio of its leading two- and one-loop logarithms agrees with that in planar $\mathcal N=4$ super Yang--Mills theory, while the opposite-helicity amplitude, which has no one-loop DPS logarithm, grows as $\ln^2(2p_T/m_A)$. In the finite virtual correction, relative to the squared tree amplitude, the highest logarithm at order $\alpha_s^2$ is $-\tfrac{7}{64}\alpha_s^2N_c^2\ln^2(2p_T/m_A)$. For $\beta_A\leq0.2$ the series reproduce the unexpanded amplitudes with relative errors of at most $6.77\times10^{-5}$ at the tested points; they provide this two-loop contribution to the hard function for soft-gluon resummation of the pair invariant-mass distribution near $m_{AA}=2m_A$ and are released with code for their numerical evaluation.
}


\begin{document}
\maketitle
\flushbottom

\input{sections/01_introduction.tex}
\input{sections/02_setup.tex}
\input{sections/03_expansion.tex}

\input{sections/04_physics.tex}
\input{sections/05_rate.tex}
\input{sections/06_conclusions.tex}

\input{sections/data_availability.tex}

\acknowledgments
I thank IFIC, University of Valencia, and Prof.\ Francisco
Campanario for providing the computing resources used in
this work.
The calculations, analysis, and code in this work were developed by the author. Claude 5.5 (Anthropic) and GPT-5.6 Pro (OpenAI) were used in a limited supporting role to assist with code debugging and optimisation and with grammatical corrections to the manuscript. The code was independently checked using the tests described in Secs.~\ref{sec:expansion} and~\ref{sec:physics} and App.~\ref{app:normalization}, including comparison with the unexpanded remainders, the renormalisation group relations, and an independent Python evaluation. These tests are included in the released package.

\appendix
\input{sections/app_constants.tex}

\input{sections/app_normalization.tex}
\input{sections/app_flavour.tex}
\input{sections/app_data.tex}

\label{end-of-appendices}
\label{start-of-references}
\bibliographystyle{JHEP}
\bibliography{references}

\end{document}

%% file: diagram_setup.tex
\IfFileExists{tikz-feynman.sty}{%
  \usepackage[compat=1.1.0]{tikz-feynman}%
  \tikzfeynmanset{warn luatex=false}%
  \newcommand{\diagramsfigure}{\input{figures/fig0_diagrams_tikz.tex}}%
}{%
  \newcommand{\diagramsfigure}{\includegraphics{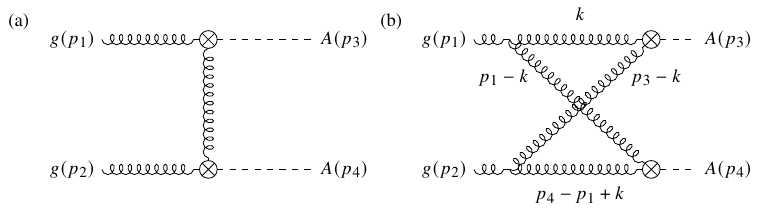}}%
}

%% file: figures/fig0_diagrams_tikz.tex
{\small
\begin{tikzpicture}
  \begin{feynman}
    \vertex (g1) at (0.9, 2.2) {\(g(p_1)\)};
    \vertex (g2) at (0.9, 0.0) {\(g(p_2)\)};
    \vertex[crossed dot] (v1) at (3.2, 2.2) {};
    \vertex[crossed dot] (v2) at (3.2, 0.0) {};
    \vertex (a3) at (5.5, 2.2) {\(A(p_3)\)};
    \vertex (a4) at (5.5, 0.0) {\(A(p_4)\)};
    \diagram* {
      (g1) -- [gluon] (v1) -- [scalar] (a3),
      (g2) -- [gluon] (v2) -- [scalar] (a4),
      (v1) -- [gluon] (v2),
    };
  \end{feynman}
  \node[anchor=north west] at (-0.3, 2.75) {(a)};

  \begin{scope}[xshift=6.3cm]
  \begin{feynman}
    \vertex (h1) at (0.9, 2.2) {\(g(p_1)\)};
    \vertex (h2) at (0.9, 0.0) {\(g(p_2)\)};
    \vertex (s1) at (2.0, 2.2);
    \vertex (s2) at (2.0, 0.0);
    \vertex[crossed dot] (e1) at (4.4, 2.2) {};
    \vertex[crossed dot] (e2) at (4.4, 0.0) {};
    \vertex (b1) at (5.7, 2.2) {\(A(p_3)\)};
    \vertex (b2) at (5.7, 0.0) {\(A(p_4)\)};
    \diagram* {
      (h1) -- [gluon] (s1) -- [gluon] (e1) -- [scalar] (b1),
      (h2) -- [gluon] (s2) -- [gluon] (e2) -- [scalar] (b2),
      (s1) -- [gluon] (e2),
      (s2) -- [gluon] (e1),
    };
    \node at (3.2, 2.62) {\(k\)};
    \node at (3.2, -0.44) {\(p_4-p_1+k\)};
    \node[anchor=east] at (2.45, 1.55) {\(p_1-k\)};
    \node[anchor=west] at (3.95, 1.55) {\(p_3-k\)};
  \end{feynman}
  \node[anchor=north west] at (-0.3, 2.75) {(b)};
  \end{scope}
\end{tikzpicture}}

%% file: metadata.tex
\newcommand{\PaperTitlePDF}{Two-loop gg to AA amplitudes near threshold and the double parton scattering singularity}
\newcommand{\PaperAuthor}{Arunima Bhattacharya}

\newcommand{\PaperEmail}{arunimab@ific.uv.es}

%% file: macros.tex
\newcommand{\as}{a_s}
\newcommand{\bA}{\beta_A}

\newcommand{\mA}{m_A}

\newcommand{\muR}{\mu_R}
\newcommand{\nf}{n_f}
\newcommand{\Nc}{N_c}
\newcommand{\Fin}{\mathrm{fin}}
\newcommand{\Ord}{\mathcal{O}}
\newcommand{\Li}{\operatorname{Li}}
\newcommand{\Real}{\operatorname{Re}}
\newcommand{\Imag}{\operatorname{Im}}
\newcommand{\dd}{\mathrm{d}}
\newcommand{\Mfull}[1]{\mathcal{M}^{(2)}_{#1}}
\newcommand{\Mflav}[2]{\mathcal{M}^{(2)}_{#1,#2}}

\providecommand{\lesssim}{\mathrel{\raisebox{-0.6ex}{$\stackrel{\textstyle<}{\sim}$}}}

%% file: numbers.tex
\newcommand{\AnglePointCount}{5}
\newcommand{\AngularReferenceChecks}{432}
\newcommand{\AngularSeriesChecks}{864}

\newcommand{\ArchiveOneLoopDev}{\ensuremath{5\times10^{-56}}}
\newcommand{\ArchiveTwoLoopDev}{\ensuremath{5\times10^{-31}}}

\newcommand{\BasisCountWord}{eight}
\newcommand{\BenchAlphaS}{0.118}
\newcommand{\BenchAs}{0.0094}
\newcommand{\BenchCos}{1/3}
\newcommand{\BenchCotBeta}{1}
\newcommand{\BenchMA}{125}
\newcommand{\BenchMt}{172.5}

\newcommand{\BenchNc}{3}
\newcommand{\BenchNf}{5}
\newcommand{\BetaPointCount}{9}

\newcommand{\BoseStatement}{are identical in all stored digits}
\newcommand{\CataniImagDigits}{15}
\newcommand{\ClosedFormBetaMax}{0.99}
\newcommand{\ClosedFormBetaMin}{0.001}
\newcommand{\ClosedFormCosMax}{0.999}
\newcommand{\ClosedFormDev}{\ensuremath{10^{-105}}}
\newcommand{\ClosedFormDigits}{120}
\newcommand{\ClosedFormPoints}{1350}
\newcommand{\ClosedFormScales}{$(125,125)$, $(125,250)$ and $(300,40)$}
\newcommand{\ClosedFormSets}{$(3,5)$, $(3,0)$, $(4,2)$, $(5,6)$ and $(2,1)$}

\newcommand{\ConstantCountWord}{thirteen}
\newcommand{\CrossoverBeta}{\ensuremath{1.7\times10^{-3}}}
\newcommand{\DPSBetaMax}{0.7}
\newcommand{\DPSBetaMin}{0.001}
\newcommand{\DPSMaxDeviation}{\ensuremath{3\times10^{-15}}}
\newcommand{\DPSNcValues}{3 and 4}
\newcommand{\DPSNfValues}{0, 2 and 5}
\newcommand{\EdgeBeta}{0.3}
\newcommand{\ExactComponentBeta}{1/10}
\newcommand{\ExactComponentCos}{1/3}
\newcommand{\ExactComponentCount}{six}

\newcommand{\ExactMaxComplete}{\ensuremath{1.3\times10^{-40}}}
\newcommand{\ExactMaxComponent}{\ensuremath{1.6\times10^{-40}}}
\newcommand{\ExactParameterSets}{three}
\newcommand{\ExactPointCount}{four}
\newcommand{\FeynmanDigits}{14}
\newcommand{\FileFactor}{64}
\newcommand{\FileFactorRoot}{8}
\newcommand{\ForwardBetaList}{$0.1$, $0.2$ and $0.3$}
\newcommand{\ForwardCosList}{$0.9$, $0.99$, $0.999$ and $-0.99$}
\newcommand{\ForwardCosMax}{0.999}
\newcommand{\ForwardSummary}{All values are below the maximum on the main grid at the same $\bA$, except for $\mathcal M_1$ at $\bA=0.1$, where the largest value, \ensuremath{1.23\times10^{-8}} at $c=0.999$, exceeds it by 21\%.}
\newcommand{\GJRatio}{-48}
\newcommand{\GridPointCount}{135}
\newcommand{\HHBetaMax}{0.7}
\newcommand{\HHBetaMin}{0.001}
\newcommand{\HHDpsMaxDev}{\ensuremath{3\times10^{-15}}}
\newcommand{\HHMtwoMax}{\ensuremath{3\times10^{-14}}}
\newcommand{\HHReproduceDigits}{40}
\newcommand{\IntroExampleBeta}{0.44}

\newcommand{\JzBetaMax}{0.5}
\newcommand{\JzBetaMin}{0.05}
\newcommand{\JzCheckAgreement}{51}
\newcommand{\JzCheckDigits}{180}
\newcommand{\JzDigits}{140}
\newcommand{\JzExponents}{$10^{-5}$, $10^{-6}$ and $10^{-7}$}
\newcommand{\JzImag}{\ensuremath{2\times10^{-4}}}
\newcommand{\JzKOneRatio}{3.4}
\newcommand{\JzMaxDev}{\ensuremath{6\times10^{-5}}}
\newcommand{\JzMaxDevWide}{\ensuremath{7\times10^{-4}}}
\newcommand{\JzSets}{$(3,5)$ and $(4,2)$}
\newcommand{\KOneBetaMax}{0.9}
\newcommand{\KOneBetaMin}{0.05}
\newcommand{\KOneClosedDev}{\ensuremath{2\times10^{-34}}}
\newcommand{\LeadCheckedCount}{24}
\newcommand{\LeadOneOneNc}{1454+462\ln2+252\ln^22-42\pi^2+693\,i\pi-693L_\mu}
\newcommand{\LeadOneOneNf}{152+84\ln2+126\,i\pi-126L_\mu}
\newcommand{\LeadOneOnePrefactor}{\frac{\pi\Nc}{9}}
\newcommand{\LeadOneThree}{-\tfrac{28}{3}\pi\Nc^2}
\newcommand{\LeadOneTwo}{-\tfrac{14}{3}\pi\Nc(11\Nc-2\nf)}
\newcommand{\LeadPowerDevLarge}{0.3}
\newcommand{\LeadTwoOne}{-\tfrac{4}{3}\pi\Nc(4\Nc-\nf)}
\newcommand{\LeadTwoTwo}{-2\pi\Nc^2}
\newcommand{\LogRelationCubic}{- \tfrac{4 \Nc}{3}}
\newcommand{\LogShareBetaLarge}{0.1}
\newcommand{\LogShareBetaSmall}{0.01}
\newcommand{\LogShareLarge}{20}
\newcommand{\LogShareSmall}{30}
\newcommand{\LoopSizeBeta}{0.1}
\newcommand{\MoneFwdBetaMax}{0.7}
\newcommand{\MoneFwdBetaMin}{0.05}
\newcommand{\MoneFwdCheckAgreement}{82}
\newcommand{\MoneFwdCheckDigits}{220}
\newcommand{\MoneFwdDevThree}{\ensuremath{4\times10^{-6}}}
\newcommand{\MoneFwdDevTwo}{\ensuremath{2\times10^{-5}}}
\newcommand{\MoneFwdDigits}{180}
\newcommand{\MoneFwdImag}{\ensuremath{2\times10^{-5}}}
\newcommand{\MoneFwdSets}{$(3,5)$ and $(4,2)$}
\newcommand{\MoneFwdWindowFactor}{four}
\newcommand{\MoneOneLoopRatio}{1.1}
\newcommand{\MoneTwoLoopRatio}{0.7}
\newcommand{\MtwoOneLoopRatio}{2.3}
\newcommand{\MtwoRelation}{- 4 \Nc}
\newcommand{\OneLoopAgreementDigits}{30}
\newcommand{\OneLoopAone}{7\pi\Nc}
\newcommand{\OneLoopAzero}{-93.900+11.484\,i}
\newcommand{\OneLoopDigits}{400}
\newcommand{\OneLoopFitBetas}{$10^{-40}$ and $10^{-50}$}
\newcommand{\OneLoopMtwo}{\tfrac{1}{2}\pi\Nc}
\newcommand{\OneLoopPowerBetaMax}{10^{-3}}
\newcommand{\OneLoopPowerBetaMin}{10^{-6}}

\newcommand{\OneLoopScanNc}{3, 4 and 5}
\newcommand{\OneLoopScanNf}{0, 2 and 5}
\newcommand{\PSLQDigits}{110}
\newcommand{\PSLQResidual}{\ensuremath{4\times10^{-107}}}
\newcommand{\ParameterSetCount}{3}

\newcommand{\PoneCancellationMax}{46.9\%}
\newcommand{\PoneCancellationMin}{42.2\%}
\newcommand{\PoneErrAtPointThree}{\ensuremath{1.10\times10^{-4}}}
\newcommand{\PoneErrLePointTwo}{\ensuremath{3.23\times10^{-6}}}
\newcommand{\PoneSlopeFour}{6.005}
\newcommand{\PoneSlopeSix}{8.005}
\newcommand{\PtwoErrAtPointThree}{\ensuremath{1.94\times10^{-3}}}
\newcommand{\PtwoErrLePointTwo}{\ensuremath{6.77\times10^{-5}}}
\newcommand{\PtwoSlopeFour}{5.955}
\newcommand{\PtwoSlopeSix}{8.034}
\newcommand{\PublishedTwoLoopMB}{204}
\newcommand{\QuadBetas}{$0.01$, $0.1$ and $0.3$}
\newcommand{\QuadNodes}{64}
\newcommand{\QuadNodesRef}{128}
\newcommand{\QuadVoneDev}{\ensuremath{4\times10^{-52}}}
\newcommand{\QuadVtwoDev}{\ensuremath{10^{-5}}}
\newcommand{\RGCoefficient}{3}
\newcommand{\RGCoefficientPublished}{192}
\newcommand{\RGSecondDev}{\ensuremath{1\times10^{-8}}}
\newcommand{\RGSets}{$(3,0,300\,\mathrm{GeV})$, $(3,5,125\,\mathrm{GeV})$ and $(4,2,150\,\mathrm{GeV})$}
\newcommand{\RGaDev}{\ensuremath{2\times10^{-7}}}
\newcommand{\RGbDev}{\ensuremath{2\times10^{-5}}}

\newcommand{\RecommendedBeta}{0.2}
\newcommand{\RenormPointsWord}{seven}

\newcommand{\ScalingFixedDev}{\ensuremath{6\times10^{-57}}}
\newcommand{\ScalingRawDev}{\ensuremath{7\times10^{-56}}}
\newcommand{\SeriesKB}{40}
\newcommand{\SeriesMassRatio}{1.02}
\newcommand{\SigmaBeta}{0.1}
\newcommand{\SigmaLO}{\ensuremath{1.29\times10^{-8}}}
\newcommand{\SizeRatio}{5061}

\newcommand{\SqrtSPointTwo}{255.2}
\newcommand{\SquareShareLarge}{4}
\newcommand{\SquareShareSmall}{21}
\newcommand{\ThresholdPercentPointThree}{4.83}
\newcommand{\ThresholdPercentPointTwo}{2.06}
\newcommand{\TimingBetas}{$1/10$, $1/5$ and $3/10$}
\newcommand{\TimingCos}{0}
\newcommand{\TimingExactMiB}{42.77}
\newcommand{\TimingFirstMax}{4.83}

\newcommand{\TimingFirstMin}{3.84}
\newcommand{\TimingMA}{125}

\newcommand{\TimingNc}{3}
\newcommand{\TimingPrecision}{80}
\newcommand{\TimingRepeats}{seven}
\newcommand{\TimingRepresentationRatio}{70.4}
\newcommand{\TimingSeriesMiB}{0.61}
\newcommand{\TimingWarmMax}{19.22}
\newcommand{\TimingWarmMedian}{19.18}
\newcommand{\TimingWarmMin}{18.99}
\newcommand{\TopMassDPSA}{1.096}
\newcommand{\TopMassDPSB}{1.145}
\newcommand{\TopMassDPSC}{1.196}
\newcommand{\TopMassMAA}{125}
\newcommand{\TopMassMAB}{150}
\newcommand{\TopMassMAC}{170}
\newcommand{\TopMassPercentMax}{2.2}
\newcommand{\TopMassRatioA}{1.006}
\newcommand{\TopMassRatioB}{1.013}
\newcommand{\TopMassRatioC}{1.022}
\newcommand{\ValAltCos}{$2/3$}
\newcommand{\ValBetaMax}{3/10}
\newcommand{\ValBetaMin}{\ensuremath{10^{-3}}}

\newcommand{\ValMaxCos}{0}
\newcommand{\ValMaxMA}{150}
\newcommand{\ValMaxMuR}{75}
\newcommand{\ValMaxNc}{4}
\newcommand{\ValSpreadWord}{eight}
\newcommand{\ValidationRows}{1080}
\newcommand{\VirtualBetaMax}{0.3}
\newcommand{\VirtualBetaMin}{\ensuremath{10^{-3}}}
\newcommand{\VoneFit}{-1508.5}
\newcommand{\VoneFormulaDev}{\ensuremath{1\times10^{-10}}}
\newcommand{\VoneMax}{\ensuremath{1.08}}
\newcommand{\VoneMin}{\ensuremath{1.06}}

\newcommand{\VoneThreshold}{115.11}
\newcommand{\VtwoAngleSpread}{\ensuremath{5\times10^{-11}}}
\newcommand{\VtwoBetaLarge}{10^{-6}}
\newcommand{\VtwoBetaSmall}{10^{-8}}
\newcommand{\VtwoLogExact}{-155.446269}
\newcommand{\VtwoLogFit}{-155.446269}
\newcommand{\VtwoLogRelDiff}{\ensuremath{3\times10^{-11}}}
\newcommand{\VtwoLogScaled}{\ensuremath{-0.0137}}
\newcommand{\VtwoMax}{\ensuremath{1.03}}
\newcommand{\VtwoMin}{\ensuremath{0.79}}
\newcommand{\VtwoPeakBeta}{0.0041}
\newcommand{\VtwoPeakEll}{-4.85}
\newcommand{\VzeroBench}{8118}
\newcommand{\VzeroTwoMu}{15421}

%% file: sections/01_introduction.tex
\section{Introduction}
\label{sec:introduction}

Precision tests of the Standard Model (SM) and searches for new particles
at the Large Hadron Collider (LHC) require theoretical predictions for
production rates and differential cross sections.
QCD radiative corrections in gluon fusion processes can be sizeable, and
calculations beyond leading order (LO) are needed to assess their effect
on the predicted rates and distributions~\cite{Baglio:2023euv}.
Loop amplitudes enter the virtual contributions to these corrections.
They can develop singularities at special kinematic configurations even
when the corresponding tree amplitudes are regular.
The location and strength of these singularities matter for numerical
evaluation and for the behaviour of differential cross sections~\cite{BDFS2012,GauntStirling2011,DixonEsterlis2016,DixonEsterlis2016Erratum}.
An expansion about such a configuration allows the singular terms to be studied explicitly, together with the regular terms needed for a quantitative prediction.

A relevant example is neutral Higgs boson pair production in
two-Higgs-doublet models (2HDM).
We consider pseudoscalar pair production in the region of 2HDM parameter space where gluon fusion through top quark loops dominates, so that loop amplitudes enter already at LO \cite{Plehn:1996wb,Plehn:1996wbErratum,Baglio:2023euv}.
For pseudoscalar pairs, $gg\to AA$, the next-to-leading order (NLO) QCD corrections were first computed in the limit of a heavy top quark~\cite{DawsonDittmaierSpira1998}, later matched to parton showers~\cite{Hespel:2014sla}, and subsequently obtained with the full dependence on the top quark mass~\cite{Baglio:2023euv}.
NLO QCD corrections in the heavy top quark limit have also been
computed for Higgs pair production with CP-violating
couplings~\cite{Grober:2017gut}.
In the benchmark considered in \cite{Baglio:2023euv}, the
full NLO cross section for $AA$ lies about $5\%$ below the
Born-improved heavy top quark approximation at LHC energies,
compared with about $12\%$ for $hH$. For this benchmark, the
comparison supports using the heavy top quark limit (HTL) as a
starting point for higher orders, with the exact mass dependence retained at LO.
The two-loop finite remainders in this limit, that is, the amplitudes after subtraction of their infrared poles~\cite{Catani:1998bh}, were obtained in~\cite{Bhattacharya:2019bmd} for the class of diagrams studied in this paper (Type~II, defined below).
For SM Higgs pairs, NLO QCD corrections are known with full top
quark mass dependence~\cite{BorowkaEtAl2016,BorowkaEtAl2016Erratum,Baglio:2020ini},
corrections in the HTL are known to N$^3$LO~\cite{deFlorianMazzitelli2013NNLO,Banerjee:2018lfq,ChenEtAl2020},
and the top-Yukawa and light-quark-induced electroweak corrections have been computed with full top quark mass dependence~\cite{Bhattacharya:2025hh}.
The status of these higher-order corrections to SM Higgs pair production, and the resulting recommendations for the total cross section and the pair invariant mass distribution, are summarised in Ref.~\cite{H:2026ksi}.
The same effective gluon interactions also contribute to the production of axion-like particle pairs, and non-resonant $gg\to aa$ production has been studied for its expected sensitivity to their interactions at the LHC~\cite{Brivio:2026alp}.

In the limit of a heavy top quark, a pseudoscalar field
$\Phi^A(x)$ couples to gluons and light quarks through the
effective Lagrangian~\cite{Chetyrkin:1998mw}
\begin{equation}
 \mathcal L^A_{\mathrm{eff}}=\Phi^A(x)\left[
 -\frac{1}{8}C_G\,O_G(x)-\frac{1}{2}C_J\,O_J(x)\right],
 \label{eq:EL}
\end{equation}
with the gluonic and light quark operators
\begin{align}
O_G(x)&=G^{a\mu\nu}\tilde{G}_{\mu\nu}^a=\epsilon_{\mu\nu\rho\sigma}
G^{a\mu\nu}G^{a\rho\sigma},
\\
G^{a\mu\nu}&=\partial^{\mu}G^{a\nu}-\partial^{\nu}G^{a\mu}+g_s f^{abc}
G_{b}^{\mu}G_{c}^{\nu},
\end{align}
where $f^{abc}$ are the structure constants of SU($\Nc$), with $\Nc=3$ in QCD, and
$\epsilon_{\mu\nu\rho\sigma}$ is the Levi-Civita tensor. The
light quark operator is the divergence of the flavour singlet
axial vector current,
\begin{equation}
O_{J}(x)=\partial_{\mu}\left(\bar{\psi}\gamma^{\mu}\gamma^{5}\psi\right) \,.
\label{eq:Fields EL}
\end{equation}
By the Adler--Bardeen theorem~\cite{Adler:1969gk,AdlerBardeen1969}, $C_G$ receives
no QCD corrections beyond one loop, whereas $C_J$ starts at second
order in the strong coupling $\as\equiv g_s^2/16\pi^2=\alpha_s/4\pi$.
The Wilson coefficients are
\begin{align}
C_{G}\left(a_{s}\right) & =-a_{s}2^{\tfrac{5}{4}}G_{F}^{\tfrac{1}{2}}\cot\beta \,,
\label{WCG}
\\
C_{J}\left(a_{s}\right) & =-\left[a_{s}C_{F}\left(\dfrac{3}{2}-3\ln\dfrac{\mu_{R}^{2}}{m_{t}^{2}}\right)
+a_{s}^{2}C_{J}^{\left(2\right)}+...\right]C_{G}\,,
\label{WCJ}
\end{align}
where $G_F$ is the Fermi constant, $m_t$ the top quark mass and
$\tan\beta$ the ratio of the vacuum expectation values of the two
Higgs doublets, in a CP-conserving 2HDM; this
angle is unrelated to the velocity
\begin{equation}
 \bA=\sqrt{1-4\mA^2/\hat s}
 \label{eq:beta_def}
\end{equation}
of the pseudoscalars in the partonic centre of mass frame, at fixed scattering angle.
$C_A=\Nc$ and $C_F=(\Nc^2-1)/(2\Nc)$ are the quadratic Casimirs of the adjoint and fundamental representations, respectively.
$\muR$ is the scale at which $\as$ is renormalised. The conventions are those of Ref.~\cite{Bhattacharya:2019bmd}. Operator mixing, ultraviolet renormalisation and the treatment of $\gamma_5$~\cite{Larin1993} follow the original calculation in Ref.~\cite{Bhattacharya:2019bmd}; the same effective theory underlies inclusive pseudoscalar production at NNLO~\cite{HarlanderKilgore2002}.

We consider the dominant gluon fusion subprocess
\begin{equation}
 g(p_1)+g(p_2)\to A(p_3)+A(p_4),
 \label{eq:amp}
\end{equation}
with $p_1^2=p_2^2=0$ for the gluons and $p_3^2=p_4^2=\mA^2$ for the pseudoscalars.
The amplitude for $gg\to AA$ separates into two classes~\cite{Bhattacharya:2019bmd}.
Type I contains a single effective gluonic insertion, producing the pseudoscalar pair through a contact interaction or an intermediate CP-even scalar.
In Type~II, each pseudoscalar is emitted from its own vertex and the two vertices are connected by gluon exchange in the $t$ or $u$ channel (Fig.~\ref{fig:diagrams}a). In the full theory, the Type~II tree diagrams correspond to two top quark triangles joined by a gluon, the one-particle reducible contributions of Refs.~\cite{DawsonDittmaierSpira1998,Baglio:2023euv}; for Higgs pairs the corresponding class was computed at two loops~\cite{Banerjee:2018lfq}. The two-loop Type~II finite remainders for $gg\to AA$ were obtained in closed form in Ref.~\cite{Bhattacharya:2019bmd}. They contain polylogarithms of weight up to four, among them $\Li_{2,2}$, and inverse powers of the Gram determinant $tu-\mA^4$, which vanishes at the pair threshold. The expressions occupy about \PublishedTwoLoopMB~MB, and large cancellations between their terms make them hard to evaluate close to threshold.

There are two reasons to study this challenging threshold region. The first is phenomenological. In the NLO predictions of Ref.~\cite{Baglio:2023euv}, for $\mA=134.7$~GeV the invariant mass distribution of the pair is larger at $m_{AA}=300$~GeV, about 30~GeV above threshold, than at 400~GeV, illustrating the relevance of the lower invariant mass region. Threshold resummation of this distribution requires at each $m_{AA}$ a hard function obtained from the infrared (IR) finite virtual amplitude~\cite{Sterman:1986aj,Catani:1989ne,deFlorian:2012za,deFlorian:2015moa,Bhattacharya:2021nsv}, and the two-loop Type~II remainders expanded here supply their contribution to it. The point $m_{AA}=300$~GeV itself corresponds to $\bA\approx\IntroExampleBeta$ and lies outside the recommended range of the present series, $0<\bA\leq\RecommendedBeta$ at the tested angles, corresponding to $m_{AA}\lesssim\SeriesMassRatio\times2\mA$. The threshold of the resummation is also a different limit, that of soft radiation at fixed $m_{AA}$, whereas the expansion here is about $m_{AA}\to2\mA$ (Sec.~\ref{sec:use}). The forward-limit tests of Sec.~\ref{sec:two-loop} are separate from both: they examine selected logarithmic coefficients of the unexpanded amplitude at larger velocities and do not extend the validated range of the series.

The second concerns the analytic structure of loop amplitudes. When the transverse momentum $p_T$ of the final state particles vanishes, a one-loop amplitude with massless internal lines can have a Landau singularity~\cite{Landau:1959fi} in which each incoming parton splits into two collinear partons that then take part in two separate $2\to1$ scatterings.
For four-jet production, the one-loop contribution in which both incident partons split is enhanced only logarithmically when the transverse momenta of the two jet pairs balance \cite{BDFS2011,BDFS2012}.
This double parton scattering (DPS) singularity is at most logarithmic in \(p_T\) for SM amplitudes; in the processes examined in Ref.~\cite{GauntStirling2011}, it occurs only when the incoming state has \(J_z=0\).
A factorised formula for its coefficient was also derived~\cite{GauntStirling2011}.
The corresponding singularity of six-gluon amplitudes in planar \(\mathcal N=4\) super Yang--Mills theory has been studied to high loop order~\cite{DixonEsterlis2016,DixonEsterlis2016Erratum}.
However, it has not been examined beyond one loop in QCD at the level of amplitudes for a $2\to2$ process with a colour singlet final state; NLO corrections to the underlying splitting mechanism are known~\cite{Diehl:2019rdh} for double parton distributions~\cite{DiehlOstermeierSchafer2012,DiehlOstermeierSchafer2012Erratum}. The availability of a complete multiloop amplitude does not by itself make the behaviour of a particular Landau region explicit: cancellations between kinematic prefactors and logarithms can obscure the singular limit. The threshold expansion below isolates this structure in an explicit two-loop QCD amplitude.
The Type~II amplitude for $gg\to AA$ contains diagrams with this
topology: each incoming gluon can split into two gluons, which
fuse pairwise into the two pseudoscalars at the effective vertices
(Fig.~\ref{fig:diagrams}b). At threshold, $p_T\to0$ at every
scattering angle, so a threshold expansion probes this
configuration directly.

In hadronic collisions, a configuration in which a parton from
each proton splits into two can enter both the single and
double parton scattering calculations. The overlap must
therefore be removed to avoid counting the same
contribution twice~\cite{DiehlSchafer2011,BDFS2012,BDFS2014,Manohar:2012pe,Gaunt:2012dd,Diehl:2019rdh,DiehlGauntSchoenwald2017,Buffing:2017mqm,Buffing:2017mqmErratum}.
DPS also receives a contribution in which only one of the
two parton pairs arises from a perturbative splitting,
while the other proton supplies two partons without such
a splitting~\cite{BDFS2011,DiehlSchafer2011,RyskinSnigirev2011,BDFS2012,BDFS2014}. This contribution has been studied for Higgs
boson pairs~\cite{GauntMaciulaSzczurek2014}.
DPS has also been observed
in same-sign $W$ pair production at the
LHC~\cite{CMS:2022pio,ATLAS:2025dps}; estimates of its effective cross section that include the contribution with one perturbative splitting are given in Ref.~\cite{BDFS2014}. The overlap between single
and double parton scattering has been re-examined using two-loop
splitting kernels~\cite{DiehlPloessl2026}. The amplitude-level singularity studied here is not itself a double parton distribution or a hadronic DPS cross section. It is the corresponding singular region of the single parton scattering loop amplitude, and therefore provides complementary information on the perturbative splitting configuration that also enters the overlap between single and double parton scattering.

In this paper, we expand the two-loop Type~II finite remainders of Ref.~\cite{Bhattacharya:2019bmd} in the velocity $\bA$ through $\bA^6$. The main results are the following:
\begin{enumerate}
 \item After the constants that arise in the expansion are expressed in a basis of \BasisCountWord{} transcendental constants, all inverse powers of $\bA$ and the odd powers through $\bA^5$ cancel, and every coefficient is even in $\cos\theta$, as required by Bose symmetry. At leading power, the angle enters only through the logarithm of the transverse momentum $p_T$ of the pseudoscalars.

 \item This logarithm is the DPS singularity. At one loop its coefficient follows the factorised formula of Ref.~\cite{GauntStirling2011} for gluon splitting and the opposite signs of the amplitudes for $g^+g^+\to A$ and $g^-g^-\to A$: with the overall constant fixed at threshold, it reproduces the velocity dependence of the exact one-loop coefficient in the forward limit. A one-loop calculation of $gg\to HH$ confirms the corresponding helicity sum for a CP-even coupling.

 \item At two loops the $J_z=0$ amplitude, with equal gluon helicities, grows as $\ln^3(2p_T/m_A)$. The ratio of its leading two and one-loop logarithms agrees with the corresponding leading-logarithmic ratio in planar $\mathcal{N}=4$ super Yang--Mills theory~\cite{DixonEsterlis2016,DixonEsterlis2016Erratum}. Forward-limit fits support the coefficient relations in Eq.\ \eqref{eq:forward-relation} at the tested velocities in $0.05\leq\beta_A\leq0.7$. The $|J_z|=2$ amplitude has no DPS logarithm at one loop but develops a $\ln^2(2p_T/m_A)$ term at two loops. Fits at the tested velocities in $0.05\leq\beta_A\leq0.5$ support a real coefficient equal to $-4N_c$ times the one-loop forward amplitude. The origin of this double logarithm remains open.

 \item At leading power, the one-loop DPS logarithm does not contribute to the finite virtual interference with the tree amplitude because the two are out of phase. At the next order, the cubic logarithm of the two-loop amplitude gives a quadratic logarithm in its interference with the tree amplitude. Combining this interference with the square of the one-loop amplitude gives
 \[V_2\supset-\frac{7}{4}\pi^2N_c^2 \ln^2\left(\frac{2p_T}{m_A}\right). \]
 Here $V_2$ is the finite virtual coefficient at two-loop order, normalised to the squared Type II tree amplitude, as defined in Eq.~\eqref{eq:virtual}.
\end{enumerate}
We use the common normalisation defined in Sec.~\ref{sec:normalization} and checked
in App.~\ref{app:normalization}. At the tested points with $0<\beta_A\leq0.2$, the series reproduce the two complete unexpanded remainders with relative errors of at most $6.77\times10^{-5}$. The series and code for their numerical evaluation are provided with the paper \cite{SupplementalMaterial}.

The paper is organised as follows.
Section~\ref{sec:setup} sets out the conventions, the normalisation of the remainders and the counting of orders. Section~\ref{sec:expansion} summarises the procedure and describes the expansion and its accuracy. Section~\ref{sec:physics} discusses the leading power terms and their relation to the DPS singularity, and Sec.~\ref{sec:rate} their consequences for the cross section. The conclusions are in Sec.~\ref{sec:conclusions}.

%% file: sections/02_setup.tex
\section{Amplitudes and conventions}
\label{sec:setup}

\subsection{Theoretical framework}

To study the threshold behaviour of $gg\to AA$ through two loops,
we use the amplitudes from Ref.~\cite{Bhattacharya:2019bmd} with the effective Lagrangian of Eq.~\eqref{eq:EL}. The labels Type~I and Type~II follow
that reference and distinguish the two classes of diagrams.
They are unrelated to the Yukawa types of 2HDMs.
Here we consider the contribution
with two $O_G$ insertions and omit the superscript
$\mathrm{II}$ from the amplitudes. Diagrams with one $O_G$
and one $O_J$ insertion first contribute at one loop.
Their coefficients are proportional to the $O_G$--$O_G$
tree coefficients, as discussed in
Sec.~\ref{sec:normalization}.

\begin{figure}[t]
 \centering
 \diagramsfigure
 \caption{(a) Type~II tree diagram in the $t$ channel; the $u$-channel diagram is obtained by exchanging the pseudoscalars. The crossed circles are the effective vertices of $O_G$. (b) One-loop Type~II diagram with loop momentum $k$. Its double parton scattering singularity lies at $k=xp_1$ and $p_3-k=(1-x)p_2$, where the four internal gluons are on shell and collinear with the incoming gluons: each incoming gluon splits into two, which fuse pairwise at the effective vertices. This topology is the crossed box of Ref.~\cite{GauntStirling2011}.}
 \label{fig:diagrams}
\end{figure}

\subsection{Kinematics}

The Mandelstam variables are
\begin{align}
 \hat s&=(p_1+p_2)^2,
 &t&=(p_1-p_3)^2,
 &u&=(p_2-p_3)^2,
 \label{eq:mandelstam}
\end{align}
with $\hat s+t+u=2\mA^2$. Following the conventions of Ref.~\cite{Bhattacharya:2019bmd}, we use
\begin{align}
 \hat s&=\mA^2\frac{(1+v)^2}{v},
 &t&=-\mA^2 y,
 &u&=-\mA^2 z,
 \label{eq:xyz-definition}
\end{align}
with $v+v^{-1}=y+z$; $v$ is the variable called $x$ in Ref.~\cite{Bhattacharya:2019bmd}, renamed here
because $x$ denotes a momentum fraction in Sec.~\ref{sec:physics}.
With $c=\cos\theta$ in the partonic centre of mass frame,
\begin{align}
 \hat s&=\frac{4\mA^2}{1-\bA^2},
 &t,u&=\mA^2-\frac{\hat s}{2}(1\mp\bA c),
 \label{eq:stu-beta}
\end{align}
Solving Eq.~\eqref{eq:stu-beta} for the variables of
Eq.~\eqref{eq:xyz-definition} and taking the branch $0<v\leq1$ gives the threshold substitution
\begin{align}
 v&=\frac{1-\bA}{1+\bA},
 &y,z&=\frac{1+\bA^2\mp2\bA c}{1-\bA^2}.
 \label{eq:threshold-map}
\end{align}
All three variables tend to one as $\bA\to0^+$ at fixed $c$, and the
expansion has to follow this correlated limit. The Gram determinant and the
transverse momentum are related by $tu-\mA^4=\hat s\,p_T^2$, with
\begin{equation}
 tu-\mA^4
 =\frac{4\mA^4\bA^2(1-c^2)}{(1-\bA^2)^2},
 \qquad
 p_T^2=\frac{\mA^2\bA^2(1-c^2)}{1-\bA^2}.
 \label{eq:gram-scaling}
\end{equation}
Both vanish at threshold at every angle, and also at fixed $\bA$ when $c\to\pm1$.

\subsection{Tensor decomposition and helicity amplitudes}

Every Type~II diagram carries two effective vertices, so the amplitude is proportional to $C_G^2$ at every order.
Extracting this factor, we decompose
\begin{equation}
 \mathcal{M}^{\mu\nu}_{ab}
 =C_G^2\,\delta_{ab}\left(T_1^{\mu\nu}\mathcal{M}_1+T_2^{\mu\nu}\mathcal{M}_2\right),
 \label{eq:tensor-decomposition}
\end{equation}
with
\begin{align}
 T_1^{\mu\nu}
 &=g^{\mu\nu}-\frac{p_1^\nu p_2^\mu}{p_1\!\cdot p_2},
 \label{eq:tensor-one}\\
 T_2^{\mu\nu}
 &=g^{\mu\nu}+\frac{1}{(p_1\!\cdot p_2)p_T^2}
 \Bigl[
 \mA^2p_2^\mu p_1^\nu
 -2(p_1\!\cdot p_3)p_2^\mu p_3^\nu
 -2(p_2\!\cdot p_3)p_3^\mu p_1^\nu
 +2(p_1\!\cdot p_2)p_3^\mu p_3^\nu
 \Bigr],
 \label{eq:tensor-two}\\
 p_T^2&=\frac{tu-\mA^4}{\hat s}.
 \label{eq:pt-definition}
\end{align}
For physical gluon polarisations, $T_1$ contributes only to equal helicities ($J_z=0$) and $T_2$ only to opposite helicities ($|J_z|=2$), each with a contraction of unit modulus.
Up to helicity-dependent overall phases, which drop out of every squared
quantity, the amplitudes for incoming gluon helicities
$(\lambda_1\lambda_2)$, with the common colour factor $\delta_{ab}$ suppressed,
are~\cite{Glover:1987nx}
\begin{equation}
 \mathcal A_{\pm\pm}=C_G^2\,\mathcal M_1,
 \qquad
 \mathcal A_{\pm\mp}=C_G^2\,\mathcal M_2\,e^{\pm2i\phi},
 \label{eq:helicity}
\end{equation}
where $\phi$ is the azimuthal angle of $\vec p_3$ about the beam axis.
Equal helicities correspond to $J_z=0$ along the beam and opposite helicities to $J_z=\pm2$; the phase of $\mathcal A_{\pm\mp}$ is fixed by rotations about the beam axis. Summing over the two configurations of each kind and over colour,
\begin{equation}
 \sum_{\mathrm{hel,col}}|\mathcal A|^2
=2(\Nc^2-1)\,C_G^4\left(|\mathcal M_1|^2+|\mathcal M_2|^2\right).
\end{equation}
The labels 1 and 2 refer to the tensors of
Eq.~\eqref{eq:tensor-decomposition}, not to the Type~I and Type~II diagram classes.

At tree level the two Type~II diagrams give
\begin{align}
 \mathcal M_1^{(0)}&=\frac{i\hat s}{2},
 \nonumber\\
 \mathcal M_2^{(0)}&=\frac{i\mA^2(v-z)(v^2+1)(vz-1)}{4vz(v^2-vz+1)}
 \nonumber\\
 &=2i\mA^2\bA^2(1-c^2)+\Ord(\bA^4),
 \label{eq:born}
\end{align}
in the variables of Eq.~\eqref{eq:xyz-definition}. We obtained Eq.~\eqref{eq:born} from the projections of the original calculation and checked it against a direct evaluation of the two Feynman diagrams with the vertex $iC_G\varepsilon_{\mu\alpha\rho\beta}k_1^\mu k_2^\rho\delta^{ab}$ that follows from Eq.~\eqref{eq:EL} (App.~\ref{app:normalization}).
The suppression of $\mathcal M_2^{(0)}$ by $\bA^2$ is the centrifugal
barrier of the $|J_z|=2$ state: two spin-zero particles can carry two units
of angular momentum along the beam only orbitally, at a cost of $\bA^2$
near threshold.

\subsection{Finite remainders and normalisation}
\label{sec:normalization}

The coefficients $\mathcal M_i$ of Eq.~\eqref{eq:tensor-decomposition}, from which $C_G^2$ has been factored out, are expanded as
\begin{equation}
 \mathcal{M}_i
 =\mathcal{M}^{(0)}_i
 +\as\mathcal{M}^{(1)}_i
 +\as^2\mathcal{M}^{(2)}_i+\cdots,
 \label{eq:perturbative-expansion}
\end{equation}
with $\as$ renormalised at $\muR$.
We use dimensional regularisation with $d=4+\epsilon$, following Ref.~\cite{Bhattacharya:2019bmd}. The finite remainders, called the finite part in Ref.~\cite{Bhattacharya:2019bmd}, are defined by the IR subtraction of Refs.~\cite{Bhattacharya:2019bmd,Catani:1998bh},
\begin{align}
 \mathcal M^{(1)}&=2I^{(1)}\mathcal M^{(0)}+\mathcal M^{(1),\Fin},
 \nonumber\\
 \mathcal M^{(2)}&=4I^{(2)}\mathcal M^{(0)}+2I^{(1)}\mathcal M^{(1)}+\mathcal M^{(2),\Fin}.
 \label{eq:catani}
\end{align}
We use the subtraction operators implemented in the original calculation, including the convention for the $H_g^{(2)}$ term specified in App.\ \ref{app:normalization}.
From here on $\mathcal{M}^{(n)}$ denotes the finite remainder
$\mathcal{M}^{(n),\mathrm{fin}}$.

At two loops, we decompose each finite remainder into powers
of the number of light quark flavours,
\begin{equation}
 \Mfull{i}=\Mflav{i}{0}+\nf\,\Mflav{i}{1}+\nf^2\,\Mflav{i}{2}.
 \label{eq:nf-decomposition}
\end{equation}
Each $\Mflav{i}{k}$ is finite, but none is the amplitude of
a separate process: the term linear in $\nf$, for example,
combines the diagrams with a closed quark loop with the terms
of the renormalisation and of the infrared subtraction that
depend on $\nf$. In the rest of the paper, the complete
coefficient means $\Mfull{i}$ at $\nf=\BenchNf$.
Furthermore, the remainders released with Ref.~\cite{Bhattacharya:2019bmd} are normalised differently at different orders: the two-loop remainders carry an overall factor $\FileFactor$ relative to the tree and one-loop amplitudes, and the one-loop remainders are those for $\muR=1$~GeV. In the normalisation of Eq.~\eqref{eq:perturbative-expansion}, the remainders are therefore
\begin{align}
 \mathcal M_i^{(1)}&=\mathcal M_{i,\mathrm{publ}}^{(1)}+2\beta_0\ln\frac{\muR^2}{\mathrm{GeV}^2}\,\mathcal M_i^{(0)},
 \nonumber\\
 \Mfull{i}&=\frac{1}{\FileFactor}\,\mathcal M_{i,\mathrm{publ}}^{(2)},
 \label{eq:normalisation}
\end{align}
with $\beta_0=(11N_c-2n_f)/3$ and masses in GeV. The first line restores the renormalisation-scale dependence absent from the released one-loop files.
The second removes the overall factor $64$ from the released two-loop files.
Appendix~\ref{app:normalization} checks both conversions against the original calculation and the renormalisation group relations.
The two-loop rescaling leaves the relative truncation errors unchanged.
Restoring the one-loop scale dependence changes the one-loop constant and the finite virtual corrections, but not the coefficient of the DPS logarithm.
All comparisons between perturbative orders below use Eq.~\eqref{eq:normalisation}.

The one-loop amplitude with one $O_G$ and one $O_J$ insertion enters at $\Ord(\as^4)$, the order of the two-loop $O_G$--$O_G$ amplitude, since $C_J=\Ord(\as)\,C_G$. It is proportional to the tree amplitude: in the normalisation of Eqs.~(3.6) and (3.7) of Ref.~\cite{Bhattacharya:2019bmd},
\begin{equation}
 \mathcal M^{(1),\Fin}_{i,GJ,\mathrm{publ}}
 = \GJRatio\,\nf\,\mathcal M_i^{(0)},
 \qquad i=1,2,
 \label{eq:gj-proportional}
\end{equation}
at every point of the physical phase space. The proportionality is consistent with the axial anomaly, which relates the divergence of the flavour singlet axial current to the gluonic pseudoscalar operator~\cite{Adler:1969gk,Bell:1969ts,Larin1993}. At $\Ord(\as^4)$ the mixed contribution therefore has the kinematic dependence of the tree amplitude and adds no logarithm of $p_T$.

\subsection{Perturbative order}
\label{sec:counting}

In Eq.~\eqref{eq:perturbative-expansion}, the superscript $(n)$ counts loops of gluons and light quarks in the effective theory; in a two-Higgs-doublet model the top quark loops are contained in the Wilson coefficients.
The perturbative order of a contribution to the cross
section also depends on the powers of $\as$ in the Wilson
coefficients.

Since $C_G\propto\as$, the Type~II tree, one-loop and
two-loop contributions to the full amplitude are of
order $\as^2$, $\as^3$ and $\as^4$, respectively.
In a 2HDM, the leading amplitude is of Type~I and
starts at $\Ord(\as)$, giving a LO cross section
of order $\as^2$. Interference between the leading
Type~I amplitude and the Type~II tree amplitude
contributes at order $\as^3$, that is at NLO.
This is the one-particle reducible contribution to
the virtual corrections in
Refs.~\cite{DawsonDittmaierSpira1998,Baglio:2023euv}.
The Type~II two-loop amplitude first enters through
its interference with the leading Type~I amplitude
at order $\as^5$, or N$^3$LO. The same counting
applies to the corresponding class of amplitudes
for Higgs pairs~\cite{Banerjee:2018lfq}.
The mixed $O_G$--$O_J$ contribution of Sec.~\ref{sec:normalization}, of the same order $\as^4$ as the two-loop $O_G$--$O_G$ amplitude, also contributes at this N$^3$LO.

When no Type~I amplitude is present, the Type~II
tree amplitude instead gives the LO cross section.
This is the case for an axion-like particle whose
only relevant interaction is its coupling to gluons
through $G\tilde G$. The one-loop and two-loop
terms in Eq.~\eqref{eq:perturbative-expansion}
then provide virtual contributions at NLO and
NNLO, respectively. The present results therefore
supply the $O_G$--$O_G$ part of the two-loop virtual
correction at NNLO for this process. A contact
interaction of dimension six containing two
pseudoscalars and two gluon fields would introduce
an additional Type~I amplitude, as in
Ref.~\cite{Brivio:2026alp}.

\subsection{Finite top quark mass effects}

The threshold expansion and the heavy top quark
approximation involve different limits. Taking
$\bA\to0$ places the pair invariant mass close to
$2\mA$, but imposes no restriction on $\mA/m_t$.
Finite top quark mass effects must therefore be
assessed separately when $A$ is the pseudoscalar of a 2HDM,
at values of $\tan\beta$ for which the top quark loop dominates.
No heavy top quark approximation is involved when
the coupling to $G\tilde G$ is specified directly
as an effective interaction, as in the axion-like
particle case above.
For single pseudoscalar production with jets, the heavy top quark effective theory and its form-factor-improved version have been compared with the full top and bottom quark mass dependence in Ref.~\cite{Bhattacharya:2026ucv}; the on-shell form factor used there is the factor $\tau_Af(\tau_A)$ of Eq.~\eqref{eq:top-dps} below.

To estimate the size of the finite top quark mass effects, we start from the Type~II tree amplitude, whose dependence on the top quark mass is known~\cite{Baglio:2023euv}. In the full theory each of the two triangles of a given channel couples a pseudoscalar to one on-shell gluon and one virtual gluon and is described by the same form factor $I_2(\tau_A,\lambda)$, with heavy top quark limit $I_2^\infty$. Each triangle therefore contributes a factor $I_2(\tau_A,\lambda)/I_2^\infty$ relative to the effective vertex, and the contribution of the channel is multiplied by
\begin{equation}
 \left[
 \frac{I_2(\tau_A,\lambda)}{I_2^\infty}
 \right]^2,
 \quad
 \tau_A=\frac{4m_t^2}{\mA^2},
 \quad
 \lambda=\frac{4m_t^2}{t}
 \quad\text{or}\quad
 \frac{4m_t^2}{u}.
 \label{eq:top-tree}
\end{equation}
Here $I_2$ is the loop function that also appears
in $A\to Z\gamma$ \cite{Baglio:2023euv}, with $\lambda$ evaluated at the virtuality of the exchanged gluon in the $t$ and $u$ channel, respectively.
At the pair threshold, $t=u=-\mA^2$, so both channels
have the same multiplicative factor relative to the heavy
top quark limit: \TopMassRatioA, \TopMassRatioB{} and \TopMassRatioC{} for $\mA=\TopMassMAA$, \TopMassMAB{} and \TopMassMAC~GeV, respectively, with $m_t=\BenchMt$~GeV. These differ from unity by at most about \TopMassPercentMax\%.

For the loop corrections, a partial estimate can
be made for the one-loop DPS logarithm. Its
coefficient involves the amplitudes for
$gg\to A$ with both gluons on shell, as discussed
in Sec.~\ref{sec:dps}. The relevant triangle form
factor is therefore obtained in the limit
$\lambda\to\infty$, rather than at the spacelike
virtuality of the exchanged gluon in the tree
diagrams:
\begin{equation}
 \frac{I_2(\tau_A,\infty)}{I_2^\infty}
 =\tau_A f(\tau_A),
 \qquad
 f(\tau)=\arcsin^2\!\left(\frac{1}{\sqrt{\tau}}\right).
 \label{eq:top-dps}
\end{equation}
This form holds for $\tau_A\geq1$, which includes the masses considered here. If the DPS factorisation applies to the full
theory, with the top quark triangles treated as
hard subgraphs, the coefficient of the one-loop
logarithm is multiplied by $[\tau_A f(\tau_A)]^2$,
one factor for each of the two subprocesses $gg\to A$;
for these masses, this is also the ratio of the full to the heavy top quark LO cross section for $gg\to A$.
Under this factorisation assumption, the one-loop DPS
logarithmic coefficient is multiplied, relative to its heavy
top quark limit, by factors of \TopMassDPSA, \TopMassDPSB{} and
\TopMassDPSC{} for the same three masses, respectively, while the velocity dependence and the helicity selection rule remain unchanged.
For these masses, the estimated finite top quark mass effect on
the one-loop DPS logarithmic coefficient is about $10$--$20\%$,
compared with at most \TopMassPercentMax\% for the Type~II tree
amplitude at threshold. The difference arises because the DPS
hard subprocesses contain two on-shell gluons, whereas each
tree triangle contains a spacelike virtual gluon.
These are changes to amplitudes or their logarithmic
coefficients, not to the full $AA$ cross section.
Neither this estimate nor the comparison of
tree amplitudes determines the finite top
quark mass corrections to the remaining
one-loop terms or to the complete two-loop
amplitude.

%% file: sections/03_expansion.tex
\section{Threshold expansion}
\label{sec:expansion}

The calculation has four steps, each checked independently. First, the finite remainders of Ref.~\cite{Bhattacharya:2019bmd} are split into their flavour coefficients and brought to the normalisation of Eq.~\eqref{eq:normalisation} (App.~\ref{app:normalization}). Second, each flavour coefficient is expanded about $\bA=0$ at fixed angle, through $\bA^4$ and, independently, through $\bA^6$ (Sec.~\ref{sec:construction}). Third, the constants that arise at the endpoints are reduced to a basis of \BasisCountWord{} transcendental constants, after which, with the numerically identified relations, the inverse and odd powers of $\bA$ cancel identically (Sec.~\ref{sec:cancellations} and App.~\ref{app:constants}). Fourth, the series are compared with the unexpanded remainders, evaluated in Mathematica at increasing working precision and, at selected points, with an independent Python code (Sec.~\ref{sec:validation}).  The software and the computing environment are described in App.~\ref{app:data}.

\subsection{Construction}
\label{sec:construction}

The remainders of Ref.~\cite{Bhattacharya:2019bmd} are functions of $v$, $y$ and $z$ (Eq.~\eqref{eq:xyz-definition}). We replace these variables by their expressions in $\bA$ and $c$ from Eq.~\eqref{eq:threshold-map}, and expand each flavour coefficient about $\bA=0$ at fixed $c$,
\begin{equation}
 \Mflav{i}{r}(\bA,c)
 =\sum_{n=-4}^{p}\sum_k
 C^{(i,r)}_{nk}\,\bA^n\ln^k\bA
 +\mathcal O_{\log}(\bA^{p+1}),
 \label{eq:series-structure}
\end{equation}
with $r=0,1,2$ and $p=4$ or 6. The coefficients $C^{(i,r)}_{nk}$ depend on $c$, $\mA^2$, $\muR^2$ and $\Nc$, and $\mathcal O_{\log}(\bA^n)$ stands for $\bA^n$ times a polynomial in $\ln\bA$. Inverse powers appear in intermediate steps because the tensor basis and the rational prefactors contain inverse Gram determinants, Eq.~\eqref{eq:gram-scaling}. They must be kept until all contributions at a given power have been combined.

The remainders contain logarithms, classical polylogarithms and the function $\Li_{2,2}$ of weight four~\cite{Frellesvig:2016ske,Vollinga:2004sn}, whose arguments approach correlated endpoints as $\bA\to0$. We expand the rational prefactors exactly and obtain the expansions of the special functions from differential equations in $\bA$, with boundary values at $\bA=0$. These equations have the iterated-integral form familiar from harmonic polylogarithms~\cite{RemiddiVermaseren2000} and from canonical differential equations for Feynman integrals~\cite{Henn2013}; the map $v=(1-\bA)/(1+\bA)$ is one of the transformations under which harmonic polylogarithms are closed~\cite{RemiddiVermaseren2000}.
The method thus expands the known remainders, not the loop integrals, which could instead be expanded near threshold by the method of regions~\cite{BenekeSmirnov1998}. Expansions of two-loop gluon fusion amplitudes in kinematic limits are also used to obtain compact representations that are fast to evaluate, for example for small transverse momentum and at high energy~\cite{Bellafronte:2022,Davies:2023} and about the top quark pair threshold~\cite{Groeber:2018}; the present expansion is about the pair production threshold of the massless remainders.
For the dilogarithm, for example, $\dd\Li_2(v)/\dd\bA=2(1-\bA^2)^{-1}\ln[2\bA/(1+\bA)]$ with $\Li_2(1)=\pi^2/6$ gives
\begin{align}
 \Li_2(v)
 &=
 \frac{\pi^2}{6}
 +2\bA\bigl[\ln(2\bA)-1\bigr]
 -\bA^2
 \nonumber\\
 &\quad
 +\bA^3
 \left[
   \frac{2}{3}\ln(2\bA)
   +\frac{1}{9}
 \right]
 -\frac{2}{3}\bA^4
 +\Ord\!\left(\bA^5\ln\bA\right).
 \label{eq:li2-threshold-example}
\end{align}
Since the prefactors start at $\bA^{-4}$, the special functions have to be expanded four orders beyond the final accuracy. The fourth and sixth order expansions were constructed independently, and all coefficients common to both agree.

\subsection{Endpoint constants and cancellations}
\label{sec:cancellations}

The boundary values at $\bA=0$ involve \ConstantCountWord{} nontrivial constants
$\mathrm{MB}[p,q,a,w]$, in addition to two elementary values.
They are defined in Eq.~\eqref{eq:mb-definition}
of App.~\ref{app:constants}, with $a,w\in\{-1,\tfrac12,1\}$.
Using the PSLQ algorithm~\cite{Ferguson:1999} at \PSLQDigits{} digits,
we identify integer relations expressing these constants in the basis
\begin{equation}
 \{\zeta_3,\ \pi^2\ln2,\ \ln^32,\ \pi^4,\ \pi^2\ln^22,\
 \ln^42,\ \Li_4(\tfrac12),\ \zeta_3\ln2\}.
 \label{eq:constant-basis}
\end{equation}
We use the conventional transcendental-weight assignment:
$\ln 2$ and $\pi$ have weight one, while $\zeta_n$ and
$\Li_n(1/2)$ have weight $n$; weights add under multiplication.
The first three constants in Eq.~\eqref{eq:constant-basis}
have weight three and the remaining five have weight four.
Constants of this type occur in alternating Euler
sums~\cite{Blumlein:2009cf}.
The numerically identified relations are listed in
Table~\ref{tab:mb-constants}; the largest residual is \PSLQResidual{}.
These relations are identified numerically and have not been proved
analytically; the cancellations below are exact once they are used.

After substituting these relations, we use the reflection and Landen
identities to express the dilogarithms with arguments $(1-c)/2$,
$(c-1)/(c+1)$ and $(c+1)/(c-1)$ in terms of
$\Li_2\bigl((1+c)/2\bigr)$.
With the overall factor $\mA^2$ extracted, the coefficient of each
power $\bA^n$ is a polynomial in
\begin{equation}
 \bigl\{\ln\bA,\ \ln(1\pm c),\
 \Li_2\bigl(\tfrac{1+c}{2}\bigr),\
 \ln\tfrac{\mA^2}{\muR^2}\bigr\}
 \cup\{\ln2,\ \pi,\ \zeta_3,\ \Li_4(\tfrac12)\},
 \label{eq:function-basis}
\end{equation}
with coefficients rational in $c$ and $\Nc$.
Treating the listed quantities as algebraically independent,
we verify symbolically that
\begin{enumerate}
 \item the coefficients of $\bA^{-4},\dots,\bA^{-1}$ vanish
 identically in every flavour coefficient;
 \item the coefficients of $\bA$, $\bA^3$ and $\bA^5$ vanish
 identically, so that through $\bA^6$ the expansion is a series
 in $\bA^2$ with logarithms;
 \item the \LeadCheckedCount{} coefficients at orders $\bA^0$,
 $\bA^2$, $\bA^4$ and $\bA^6$ across the six flavour components
 are even in $c$, so that Bose symmetry holds order by order.
\end{enumerate}
The cancellations and the reductions of the constants were confirmed independently in Mathematica.

The cancellations have a simple origin. Exchanging the pseudoscalars
exchanges $t$ and $u$, so the remainders depend on the kinematics through
invariants that are symmetric under this exchange,
\begin{equation}
 \hat s=\frac{4\mA^2}{1-\bA^2},\qquad
 (t-u)^2=\hat s^2\bA^2c^2,\qquad
 tu-\mA^4=\hat s\,p_T^2,
 \label{eq:symmetric-invariants}
\end{equation}
with $p_T^2$ given in Eq.~\eqref{eq:gram-scaling}. The difference
$t-u=\hat s\bA c$, which is odd in $\bA$, enters only through its square.
Each invariant depends on $\bA$ only through $\bA^2$ and on $c$ only
through $c^2$, so a function analytic in them at threshold is a series in
$\bA^2$ and even in $c$. Odd or inverse powers of $\bA$ would therefore signal
non-analytic behaviour at threshold. There is no threshold for a pair of
massive internal particles at $\hat s=4\mA^2$, since the loops contain no
massive particles, and the colourless final state has no QCD Coulomb
singularity. These observations alone do not exclude other Landau
singularities: the singularity at $p_T\to0$ discussed in Sec.~\ref{sec:dps}
is one, and at one loop it is logarithmic~\cite{BDFS2011,BDFS2012,GauntStirling2011}. The Gram
determinant $tu-\mA^4$ vanishes at the same point, but its inverse powers come
from the tensor basis and the rational prefactors; they are not singularities
of the amplitude. In the present calculation, the explicit expansion shows
that the inverse powers cancel and that the remaining singular behaviour is
logarithmic in $p_T$.
A remainder of the form $\sum_k a_k\ln^k p_T^2$, with coefficients $a_k$
analytic in the invariants of Eq.~\eqref{eq:symmetric-invariants} at
threshold, is therefore a series in $\bA^2$ with logarithms and has no
inverse powers. This accounts for the form of the cancellations found above.
At two loops it is not an independent proof, because there the logarithmic
nature of the singularity is established only by the calculation itself.

Table~\ref{tab:leading-threshold-inventory} lists the leading terms. Since
the odd powers vanish, the first omitted term is expected to be
$\Ord_{\log}(\bA^{p+2})$ rather than the $\Ord_{\log}(\bA^{p+1})$ of
Eq.~\eqref{eq:series-structure}. The truncation errors should therefore
scale as $\bA^6$ for $p=4$ and as $\bA^8$ for $p=6$, up to logarithmic
factors.

\input{tables/leading_terms.tex}

The coefficients of $\nf^2$ in Eq.~\eqref{eq:nf-decomposition}
have the simplest threshold behaviour:
\begin{align}
 \Mflav{1}{2}(\bA,c)&=K_{\nf}+\Ord(\bA^2),
 \label{eq:nf2-leading-one}\\
 \Mflav{2}{2}(\bA,c)&=\bA^2(1-c^2)K_{\nf}+\Ord(\bA^4),
 \label{eq:nf2-leading-two}
\end{align}
where $K_{\nf}$ is the angle-independent threshold value of
$\Mflav{1}{2}$, with the overall factor $\nf^2$ excluded.
The $\nf^2$ component is the only flavour component that follows the tree-level pattern $\mathcal M_2\propto\bA^2(1-c^2)$ of Eq.~\eqref{eq:born}, with the same ratio to the tree amplitude in both tensor structures. The complete expressions through $\bA^6$ are released as Wolfram Language files (App.~\ref{app:data}).

\subsection{Accuracy}
\label{sec:validation}

We compared the series with the unexpanded remainders on
the grid in Table~\ref{tab:validation-grid}, which contains
\BetaPointCount{} values of $\bA$ between \ValBetaMin{}
and $\ValBetaMax$, \AnglePointCount{} scattering angles
and \ParameterSetCount{} combinations of mass,
renormalisation scale and number of colours.
At each of the \GridPointCount{} points, the comparison
included the six flavour coefficients and the two complete
coefficients, giving \ValidationRows{} comparisons in total.
Each unexpanded remainder was evaluated using an adaptive sequence
of working precisions: 80, 120, 160, 220 and 300 digits.
Evaluation stopped at the first pair of successive precisions
that agreed to a relative difference below $10^{-25}$,
with at least 30 significant digits in both. All reference values
on the reported validation grid were accepted at the 80/120-digit pair.
The parameter set with $\Nc\neq3$ was
included to test the colour dependence; all physical
results use $\Nc=\BenchNc$.
The relative truncation error is defined as
\begin{equation}
 E_i^{(p)}
 =
 \frac{\left|\mathcal M_i^{[p]}-\mathcal M_i\right|}
      {\left|\mathcal M_i\right|},
 \qquad p=4,6,
 \label{eq:relative-error}
\end{equation}
where $\mathcal M_i$ is the unexpanded remainder and $\mathcal M_i^{[p]}$ its series through $\bA^p$.

\begin{figure*}[!htb]
 \centering
 \includegraphics[width=0.98\textwidth]{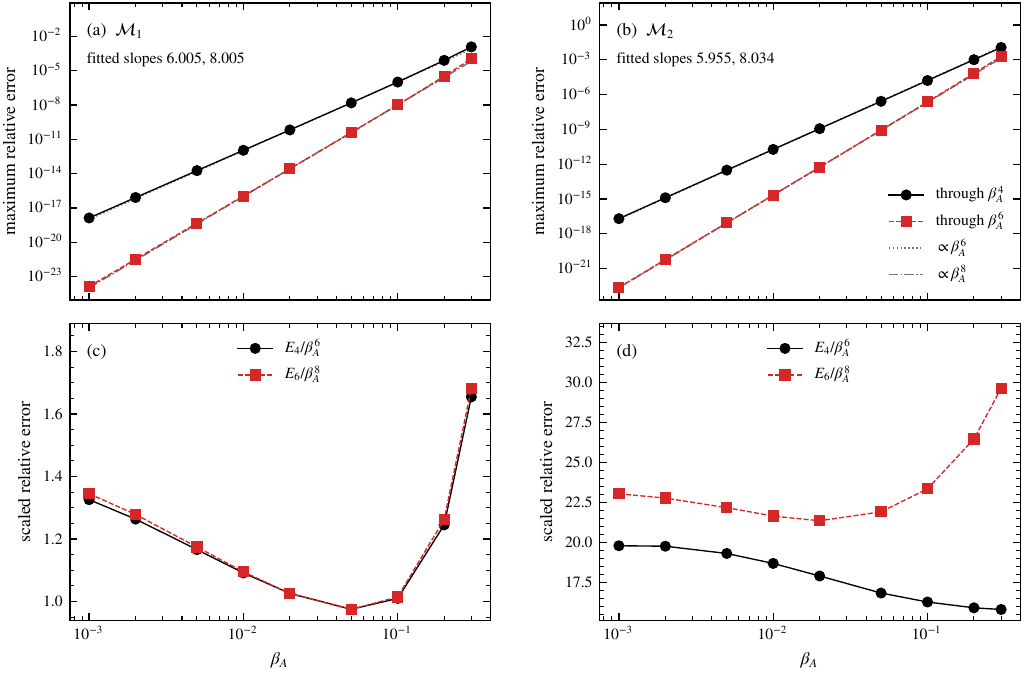}
 \caption{Relative truncation errors, Eq.~\eqref{eq:relative-error}, of the complete two-loop coefficients at $\nf=\BenchNf$: $\mathcal M_1$ (left) and $\mathcal M_2$ (right). Upper panels: maximum over the angles and parameter sets of Table~\ref{tab:validation-grid} for the series through $\bA^4$ (circles) and through $\bA^6$ (squares), with lines proportional to $\bA^6$ and $\bA^8$. Lower panels: the same maxima divided by $\bA^6$ and $\bA^8$.}
 \label{fig:validation}
\end{figure*}
\input{tables/validation_grid.tex}
\input{tables/accuracy.tex}
\input{tables/forward_angles.tex}
Figure~\ref{fig:validation} shows these errors for the complete coefficients
at $\nf=\BenchNf$. The slopes of the curves, discussed below, test the
scaling $\bA^{p+2}$ expected from the cancellation of the odd powers.
Because each curve is a maximum over the points of
Table~\ref{tab:validation-grid}, a change of the point that gives the
maximum as $\bA$ varies would distort these slopes. For the series through
$\bA^6$ this does not occur: for both coefficients and at every velocity
the maximum comes from the same point, $c=\ValMaxCos$ with $\mA=\ValMaxMA$~GeV,
$\muR=\ValMaxMuR$~GeV and $\Nc=\ValMaxNc$, so the curves follow the error of a single point.
For the series through $\bA^4$ the same holds for $\mathcal M_1$, while for $\mathcal M_2$ the maximum alternates between two parameter sets at $|c|=$\ValAltCos.
The maximum is also representative of the grid: at each velocity and
for both orders the errors at all other points lie within a factor of \ValSpreadWord{}
of it.

For the series through $\bA^4$, the fitted slopes of the logarithm of the maximum error against $\ln\bA$ over the plotted range are \PoneSlopeFour{}
for $\mathcal M_1$ and \PtwoSlopeFour{} for $\mathcal M_2$; for the series through $\bA^6$ they are \PoneSlopeSix{} and \PtwoSlopeSix.
The expected values are 6 and 8 (Sec.~\ref{sec:cancellations}), while a surviving odd power would give values close to 5 and 7. The small deviations from 6 and 8 are consistent with the powers of $\ln\bA$ and the higher terms omitted from the series.

In the lower panels, exact power behaviour would give horizontal lines.
The scaled errors vary slowly, because both the omitted terms and the remainders that normalise the errors contain powers of $\ln\bA$, and over the tested range they change by less than a factor of two.
An incorrect coefficient at a retained order would instead make them grow as an inverse
power of $\bA$ as $\bA$ decreases. Towards $\bA=\EdgeBeta$ the scaled errors of the series through $\bA^6$ rise for both coefficients, as the terms beyond the first omitted order begin to contribute.

Table~\ref{tab:AA-threshold-accuracy} lists the maximum errors at selected velocities. For $\bA\leq\RecommendedBeta$, the series through $\bA^6$ reproduces $\mathcal M_1$ and $\mathcal M_2$ with relative errors of at most \PoneErrLePointTwo{} and \PtwoErrLePointTwo{}; at $\bA=\EdgeBeta$ the errors are \PoneErrAtPointThree{} and \PtwoErrAtPointThree{}.
Bose symmetry was checked by comparing the values at $c$ and $-c$: they \BoseStatement{} in \AngularReferenceChecks{} comparisons of the unexpanded coefficients and \AngularSeriesChecks{} of the series.
The reference values were computed in Mathematica from the expressions of Ref.~\cite{Bhattacharya:2019bmd}. To check their evaluation independently, we wrote a separate Python code with mpmath~\cite{mpmath}. It reads the published files with its own parser and computes $\Li_{2,2}$ from its defining double sum, or from a one-dimensional integral representation where the sum converges slowly. At \ExactPointCount{} points covering the \ExactParameterSets{} parameter sets, the two complete coefficients agree with the reference values to a relative difference of at most \ExactMaxComplete{}, the precision to which the reference values are stored. The \ExactComponentCount{} flavour coefficients, compared at $\bA=\ExactComponentBeta$ and $c=\ExactComponentCos$, agree to \ExactMaxComponent{}. At $c=0$, where $t=u$, the published expressions have removable singularities and are evaluated as limits.

The expansion is made at fixed angle, but $p_T$ vanishes as $c\to\pm1$ at fixed $\bA$, so the series must also be tested close to the beam axis. We therefore compared it with the unexpanded remainders at $c=$ \ForwardCosList{} for $\bA=$ \ForwardBetaList{}. Table~\ref{tab:forward-angles} gives the relative errors of the series through $\bA^6$. \ForwardSummary{} The series thus remains accurate close to the beam, consistent with its coefficients having no poles at $c=\pm1$, only logarithms of $1\pm c$ (Sec.~\ref{sec:leading}).

%% file: tables/leading_terms.tex
\begin{table}[t]
\centering

\begin{tabular}{lccc}
\toprule
& inverse powers & leading & powers \\
quantity & in construction & power & of $\mathcal L$ \\
\midrule
$\Mflav{1}{0}$ & $\bA^{-2}$, $\bA^{-1}$ & $\bA^{0}$ & $0,1,2,3$ \\
$\Mflav{1}{1}$ & -- & $\bA^{0}$ & $0,1,2$ \\
$\Mflav{1}{2}$ & -- & $\bA^{0}$ & $0$ \\
$\Mflav{2}{0}$ & $\bA^{-4}$, $\bA^{-3}$, $\bA^{-2}$, $\bA^{-1}$ & $\bA^{0}$ & $0,1,2$ \\
$\Mflav{2}{1}$ & $\bA^{-1}$ & $\bA^{0}$ & $0,1$ \\
$\Mflav{2}{2}$ & -- & $\bA^{2}$ & $0$ \\
\bottomrule
\end{tabular}
\caption{Leading terms of the threshold expansion of the two-loop Type II remainders. The second column lists the inverse powers with nonzero coefficients in the construction files; they cancel identically after the reductions of Sec.~\ref{sec:cancellations}, as do all odd powers. The last column gives the powers of $\mathcal L$, Eq.~\eqref{eq:calL}, at leading power.}
\label{tab:leading-threshold-inventory}
\end{table}

%% file: tables/validation_grid.tex
\begin{table}[t]
\centering

\begin{tabular}{ll}
\toprule
quantity & values \\
\midrule
$\bA$ & $1/1000$, $1/500$, $1/200$, \\
 & $1/100$, $1/50$, $1/20$, \\
 & $1/10$, $1/5$, $3/10$ \\
$c$ & $-2/3$, $-1/3$, $0$, $1/3$, $2/3$ \\
$(\mA/\mathrm{GeV},\,\muR/\mathrm{GeV},\,\Nc)$ & $(125,125,3)$, \\
 & $(125,250,3)$, \\
 & $(150,75,4)$ \\
\bottomrule
\end{tabular}
\caption{Validation grid. At each point the six flavour coefficients and the two complete $\nf=\BenchNf$ coefficients are compared.}
\label{tab:validation-grid}
\end{table}

%% file: tables/accuracy.tex
\begin{table}[!htb]
\centering

\small
\setlength{\tabcolsep}{3pt}

\begin{tabular}{ccccc}
\toprule
& \multicolumn{2}{c}{$\mathcal{M}_1$ through} & \multicolumn{2}{c}{$\mathcal{M}_2$ through} \\
$\beta_A$ & $\beta_A^4$ & $\beta_A^6$ & $\beta_A^4$ & $\beta_A^6$ \\
\midrule
0.01 & $1.09\times 10^{-12}$ & $1.09\times 10^{-16}$ & $1.87\times 10^{-11}$ & $2.16\times 10^{-15}$ \\
0.05 & $1.52\times 10^{-8}$ & $3.81\times 10^{-11}$ & $2.63\times 10^{-7}$ & $8.55\times 10^{-10}$ \\
0.10 & $1.01\times 10^{-6}$ & $1.01\times 10^{-8}$ & $1.63\times 10^{-5}$ & $2.33\times 10^{-7}$ \\
0.20 & $7.97\times 10^{-5}$ & $3.23\times 10^{-6}$ & $1.02\times 10^{-3}$ & $6.77\times 10^{-5}$ \\
0.30 & $1.21\times 10^{-3}$ & $1.10\times 10^{-4}$ & $1.15\times 10^{-2}$ & $1.94\times 10^{-3}$ \\
\bottomrule
\end{tabular}
\caption{Maximum relative errors of the complete Type II coefficients on the validation grid. At each value of $\beta_A$, the maximum is taken over the \AnglePointCount{} angles and \ParameterSetCount{} parameter sets of Table~\ref{tab:validation-grid}.}
\label{tab:AA-threshold-accuracy}
\end{table}

%% file: tables/forward_angles.tex
\begin{table}[t]
\centering

\setlength{\tabcolsep}{3pt}
\begin{tabular}{llccc}
\toprule
& $c$ & $\bA=0.1$ & $\bA=0.2$ & $\bA=0.3$ \\
\midrule
$\mathcal M_1$ & $0.9$ & $6.36\times10^{-9}$ & $1.11\times10^{-6}$ & $1.78\times10^{-5}$ \\
$\mathcal M_1$ & $0.99$ & $1.06\times10^{-8}$ & $2.43\times10^{-6}$ & $5.76\times10^{-5}$ \\
$\mathcal M_1$ & $0.999$ & $1.23\times10^{-8}$ & $2.96\times10^{-6}$ & $7.18\times10^{-5}$ \\
$\mathcal M_2$ & $0.9$ & $1.39\times10^{-7}$ & $3.08\times10^{-5}$ & $6.68\times10^{-4}$ \\
$\mathcal M_2$ & $0.99$ & $6.68\times10^{-8}$ & $1.47\times10^{-5}$ & $2.88\times10^{-4}$ \\
$\mathcal M_2$ & $0.999$ & $6.21\times10^{-8}$ & $1.65\times10^{-5}$ & $3.74\times10^{-4}$ \\
$\mathcal M_1$ & grid max. & $1.01\times10^{-8}$ & $3.23\times10^{-6}$ & $1.10\times10^{-4}$ \\
$\mathcal M_2$ & grid max. & $2.33\times10^{-7}$ & $6.77\times10^{-5}$ & $1.94\times10^{-3}$ \\
\bottomrule
\end{tabular}
\caption{Relative truncation errors $E_i^{(6)}$ of the complete $\nf=\BenchNf$ coefficients towards the beam, at $\mA=\muR=\BenchMA\,\mathrm{GeV}$ and $\Nc=\BenchNc$. The last rows give the largest value on the grid of Table~\ref{tab:validation-grid}. The values at $c=-0.99$ equal those at $-c$.}
\label{tab:forward-angles}
\end{table}

%% file: sections/04_physics.tex
\section{Leading power and the double parton scattering singularity}
\label{sec:physics}

\subsection{Leading power}
\label{sec:leading}

To determine the leading threshold behaviour of the
two-loop remainders, we first collect the terms of
order $\bA^0$, including all powers of $\ln\bA$
that multiply them. The notation $\big|_{\bA^0}$ used below
denotes this part of the expansion.

In the expanded remainders, the logarithms of $\bA$ and of the angular
variables appear at leading power in the combination
$\ln\bA+\tfrac12\ln(1-c^2)$. We therefore introduce
\begin{align}
 \mathcal L
 &=\ln\bA+\tfrac12\ln(1-c^2)+\ln2+\frac{i\pi}{2}
 \nonumber\\
 &=\tfrac12\ln\bigl[-4\bA^2(1-c^2)\bigr],
 \label{eq:calL}
\end{align}
where the logarithm of the negative argument is taken on the branch
$\ln(-X)=\ln X+i\pi$ for $X>0$, which is the one followed by the remainders
in the physical region.

The meaning of $\mathcal L$ follows from Eq.~\eqref{eq:gram-scaling}.
Substituting the expressions for $tu-\mA^4$ and $p_T^2$ gives
\begin{align}
 \nonumber
 \mathcal L
 &=\tfrac12\ln\frac{\mA^4-tu}{\mA^4}
   +\ln(1-\bA^2)
 \\
 &=\ln\frac{2p_T}{\mA}+\frac{i\pi}{2}
   +\tfrac12\ln(1-\bA^2).
 \label{eq:calL-pT}
\end{align}
The last term is of order $\bA^2$. Near threshold, $\mathcal L$ is therefore
the logarithm of the transverse momentum of the pseudoscalars, up to the
constant phase $i\pi/2$ and power corrections. Away from threshold the
natural variable is $\mathcal G=\tfrac12\ln[(\mA^4-tu)/\mA^4]=\mathcal L-\ln(1-\bA^2)$,
used in Sec.~\ref{sec:two-loop}.

When the leading power terms are rewritten in terms of $\mathcal L$, no
separate $\ln(1\pm c)$ or $\Li_2\bigl((1+c)/2\bigr)$ remains, and the
rational dependence on $c$ cancels. All dependence on the scattering angle
is then contained in $\mathcal L$, and the coefficients of its powers are
independent of $c$. For general $\Nc$ and $\nf$,
\begin{align}
 \Mfull{1}\big|_{\bA^0}
 &=\mA^2\sum_{k=0}^{3}
   \widehat A_{1k}\,\mathcal L^k,
 \label{eq:leading-power-M1}
 \\
 \Mfull{2}\big|_{\bA^0}
 &=\mA^2\sum_{k=0}^{2}
   \widehat A_{2k}\,\mathcal L^k.
 \label{eq:leading-power}
\end{align}
The highest logarithmic coefficients are
\begin{align}
 \widehat A_{13}&=\LeadOneThree,
 &
 \widehat A_{12}&=\LeadOneTwo,
 \nonumber\\
 \widehat A_{22}&=\LeadTwoTwo,
 &
 \widehat A_{21}&=\LeadTwoOne,
 \label{eq:leading-coefficients}
\end{align}
which do not depend on $\muR$, and
\begin{equation}
 \widehat A_{11}=\LeadOneOnePrefactor\bigl(\Nc\,a_{\mathrm{A}}-\nf\,a_{\mathrm{F}}\bigr),
 \label{eq:M1-L1}
\end{equation}
where
\begin{align}
 a_{\mathrm{A}}&=\LeadOneOneNc,
 \nonumber\\
 a_{\mathrm{F}}&=\LeadOneOneNf,
 \nonumber
\end{align}
and $L_\mu=\ln(\mA^2/\muR^2)$. Table~\ref{tab:leading-log-benchmark} gives all coefficients for $\Nc=\BenchNc$, $\nf=\BenchNf$ and $\muR=\mA$. At $\bA^2$, $\bA^4$ and $\bA^6$ the same substitution again removes every $\ln(1\pm c)$ and $\Li_2$; the remaining dependence on the angle is rational, even in $c$ and free of poles at $c=\pm1$.
\input{tables/leading_logs.tex}

\begin{figure*}[!htb]
 \centering
 \includegraphics[width=0.94\textwidth]{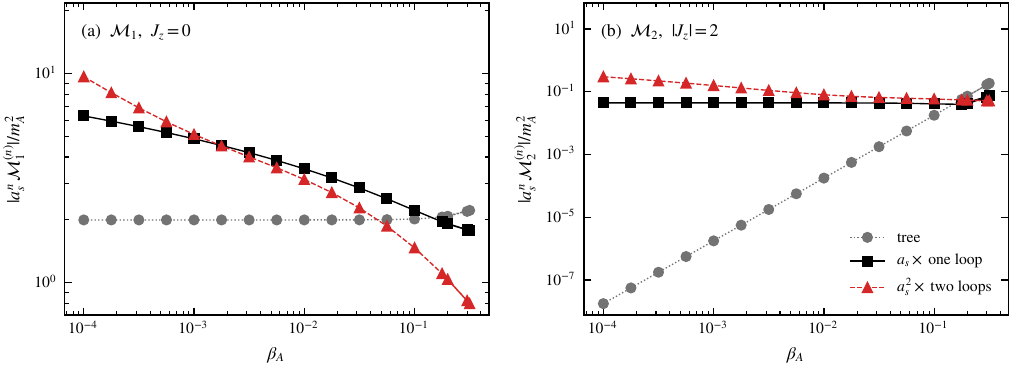}
 \caption{Tree-level, one-loop and two-loop terms of the Type~II amplitude, $|\as^n\mathcal M_i^{(n)}|/\mA^2$ with $\alpha_s=\BenchAlphaS$, at $c=\BenchCos$, $\mA=\muR=\BenchMA\,\mathrm{GeV}$, $\Nc=\BenchNc$ and $\nf=\BenchNf$. Tree level: Eq.~\eqref{eq:born}; one loop: remainder of Ref.~\cite{Bhattacharya:2019bmd} with the scale dependence of Eq.~\eqref{eq:normalisation} restored; two loops: series through $\bA^6$. $\mathcal M_1$ and $\mathcal M_2$ are the equal and opposite helicity amplitudes, Eq.~\eqref{eq:helicity}.}
 \label{fig:amplitudes}
\end{figure*}

The exact one-loop remainders already contain $\mathcal L$, since they depend on
\begin{align*}
 &\ln(y-v)+\ln(z-v) \text{ and }\\
 &(y-v)(z-v)=yz-1=(tu-\mA^4)/\mA^4
\end{align*}
by Eq.~\eqref{eq:xyz-definition}. The two-loop coefficients have no finite limit as $\bA\to0$; they grow as $\mathcal L^3$ and $\mathcal L^2$ (Fig.~\ref{fig:amplitudes}).

\subsection{One loop}
\label{sec:one-loop}

Fitting the one-loop remainders of Ref.~\cite{Bhattacharya:2019bmd} at $\bA=$ \OneLoopFitBetas{} with \OneLoopDigits-digit arithmetic, for $\Nc=$ \OneLoopScanNc, $\nf=$ \OneLoopScanNf{}, two values of $\muR$ and several angles, we find to \OneLoopAgreementDigits{} digits
\begin{align}
 \mathcal M_1^{(1)}&=\mA^2\bigl[\OneLoopAone\,\mathcal L+\widehat A^{(1)}_{10}\bigr]+\Ord_{\log}(\bA^2),
 \nonumber\\
 \mathcal M_2^{(1)}&=\OneLoopMtwo\,\mA^2+\Ord_{\log}(\bA^2),
 \label{eq:one-loop-threshold}
\end{align}
with $\widehat A^{(1)}_{10}=\OneLoopAzero$ for $\Nc=\BenchNc$, $\nf=\BenchNf$ and $\muR=\mA$, in the normalisation of Eq.~\eqref{eq:normalisation}. Evaluations at $\OneLoopPowerBetaMin\leq\bA\leq\OneLoopPowerBetaMax$ show that the first corrections are of order $\bA^2\ln\bA$ for $\mathcal M_1^{(1)}$ and $\bA^2$ for $\mathcal M_2^{(1)}$. The constant does not depend on the angle, and by Eq.~\eqref{eq:normalisation}, its scale dependence is $\partial\widehat A^{(1)}_{10}/\partial\ln\muR^2=4i\beta_0$. The coefficient of $\mathcal L$ and the limit of $\mathcal M_2^{(1)}$ depend neither on $\nf$ nor on $\muR$. Together with Eqs.~\eqref{eq:born}, \eqref{eq:leading-power-M1} and \eqref{eq:leading-power}, the behaviour at threshold is summarised in Table~\ref{tab:orders}.
\begin{table}[!htb]
\centering
 \begin{tabular}{lccc}
\toprule
  & tree level & one loop & two loops \\
  \midrule
  $\mathcal M_1$ ($J_z=0$) & $2i\mA^2$ & $\mathcal L$ & $\mathcal L^3$\\
  $\mathcal M_2$ ($|J_z|=2$) & $2i\mA^2\bA^2(1-c^2)$ & $\mathcal L^0$ & $\mathcal L^2$ \\
\bottomrule
\end{tabular}
\caption{Behaviour of the Type~II coefficients at threshold, $\bA\to0$ at fixed angle: tree-level values, and the highest power of $\mathcal L$ at one and two loops.}
\label{tab:orders}
\end{table}
The $|J_z|=2$ amplitude is suppressed near threshold only at tree level. According to Eq.~\eqref{eq:helicity}, $\mathcal A_{\pm\mp}$ retains the azimuthal factor $e^{\pm2i\phi}$ as $|\vec p_3|\to0$. The loop amplitude therefore has a direction-dependent limit and is not analytic in $\vec p_3$ at the origin. At one loop $\mathcal M_2$ approaches a nonzero constant without a DPS logarithm; at two loops it also acquires the logarithm discussed in Sec.~\ref{sec:two-loop}.

\subsection{The double parton scattering singularity}
\label{sec:dps}

\paragraph{Origin of the logarithm.}

The logarithm $\mathcal L$ in Eq.~\eqref{eq:one-loop-threshold} can be
understood from the one-loop Type~II diagrams near $p_T=0$, shown in
Fig.~\ref{fig:diagrams}(b). As $p_T\to0$, all four internal gluons can be
on shell simultaneously, two of them collinear to each incoming gluon. This
configuration is the pinch that produces the DPS singularity of the crossed
box~\cite{GauntStirling2011}.
The singular contribution can be expressed as
an integral over transverse momentum of the
product of the light cone wavefunctions of the
two incoming gluons and the tree amplitudes of
the two hard subprocesses~\cite{GauntStirling2011}.
The wavefunctions describe the splitting of each
incoming parton. The hard subprocess amplitudes
are evaluated with collinear on-shell incoming
partons and zero transverse momentum. Thus, the
splittings determine the strength of the
singularity, while the hard subprocesses enter
its coefficient. In the present case, the
splittings are $g\to gg$ and the subprocesses
are $gg\to A$ through $O_G$.

For a gluon splitting into a collinear pair,
the daughter helicities add to $0$ or $\pm2$,
whereas the parent helicity is $\pm1$.
The numerator of the splitting vertex therefore
vanishes in the exactly collinear limit; for the
allowed tree helicity configurations, it
is linear in the relative transverse momentum.
The corresponding light cone wavefunction behaves as the inverse of that momentum after including the propagator.
The transverse momentum integral is consequently at most logarithmically singular, as in the massless quark examples
of Ref.~\cite{GauntStirling2011}.

\paragraph{Helicity selection and wavefunction factors.}

The logarithm is present only for certain helicities of the partons in the
singular configuration. For the fermion loops of
Ref.~\cite{GauntStirling2011}, helicity conservation forces the four
intermediate partons into $J_z=0$; for gluon loops only conservation of
$J_z$ between the incoming state and the intermediate partons
remains~\cite{DixonEsterlis2016,DixonEsterlis2016Erratum}. Here the effective vertices restore the
restriction. The amplitude for $gg\to A$ vanishes unless the two gluons
have equal helicities, and the gluons of each fusing pair move in opposite
beam directions, so each pair has $J_z=0$. A DPS logarithm is therefore
possible only for incoming gluons with $J_z=0$, which is why
$\mathcal M_1$ has a logarithm at one loop and $\mathcal M_2$ does not.

For a positive-helicity gluon splitting into gluons $a$ and $b$ with
longitudinal momentum fractions $\xi$ and $1-\xi$, the light cone wavefunction is,
for each pair of daughter helicities, a function of $\xi$ times a phase and a power
of the relative transverse momentum; with the coupling, colour and transverse momentum
factors removed, the functions of $\xi$ are
\begin{align}
 w_{++}(\xi)&=\frac{1}{\sqrt{\xi(1-\xi)}},
 \nonumber\\
 w_{+-}(\xi)&=\frac{\xi^2}{\sqrt{\xi(1-\xi)}},
\nonumber\\
 w_{-+}(\xi)&=\frac{(1-\xi)^2}{\sqrt{\xi(1-\xi)}},
 \label{eq:splitting-weights}
\end{align}
where the subscripts give $(h_a,h_b)$~\cite{Mangano:1990by,Dixon:1996wi}.
The sum of their squares, $[1+\xi^4+(1-\xi)^4]/[\xi(1-\xi)]$, is the $\xi$
dependence of the $g\to gg$ splitting function. At the DPS point, the first
pseudoscalar receives a fraction $x$ of the first incoming momentum and,
since the two pseudoscalars have equal masses, a fraction $1-x$ of the
second~\cite{GauntStirling2011}; in Fig.~\ref{fig:diagrams}(b),
$k=xp_1$ and $p_3-k=(1-x)p_2$.

Since $A$ is odd under parity, which reverses all helicities, the
amplitudes for $g^+g^+\to A$ and $g^-g^-\to A$ have opposite
signs~\cite{Dixon:1996wi}. For two incoming positive-helicity gluons, with
the first helicity of each daughter pair referring to the gluon that goes
to the first pseudoscalar, each vertex requires equal helicities, so both
splittings produce the same pair: $(+,+)$, $(+,-)$ or $(-,+)$. In the mixed
cases one vertex receives two negative-helicity gluons, and these
contributions enter with the opposite sign. The factorised form of
Ref.~\cite{GauntStirling2011} then gives
\begin{align}
 F_A(x)
 &=w_{++}(x)w_{++}(1-x)
 -w_{+-}(x)w_{+-}(1-x)
       -w_{-+}(x)w_{-+}(1-x)
 \nonumber\\
 &=\frac{1}{x(1-x)}-2x(1-x),
 \label{eq:FA}
\end{align}
since each mixed product equals $x(1-x)$. For a CP-even coupling to $G^2$
the two signs are equal and the mixed terms add:
\begin{equation}
 F_S(x)=\frac{1}{x(1-x)}+2x(1-x).
 \label{eq:FS}
\end{equation}
We call $F_A$ and $F_S$ helicity sums: they are sums over the daughter helicities of
products of two wavefunction factors, overlaps of amplitudes rather than probabilities, and
the relative signs in $F_A$ come from the interference of the helicity configurations.
These helicity sums differ from those of the massless quark boxes for
$gg\to HH$ in Ref.~\cite{GauntStirling2011}, where internal quarks of equal
helicity carry complementary momentum fractions and the sum is
proportional to $x(1-x)$.

\paragraph{Normalisation and the forward limit.}

At threshold, both pseudoscalars are at rest
and $x=\tfrac12$, so that
$F_A(\tfrac12)=\tfrac72$ and
$F_S(\tfrac12)=\tfrac92$.
By Eq.~\eqref{eq:calL}, the coefficient of
$\ln p_T^2$ is half the coefficient of
$\mathcal L$. The one-loop threshold result
in Eq.~\eqref{eq:one-loop-threshold}
therefore gives
\[
 \frac{\OneLoopAone\,\mA^2}{2}
 =\frac72\pi\Nc\mA^2
 =\pi\Nc\mA^2F_A\!\left(\frac12\right).
\]
Matching at threshold thus fixes the overall
factor multiplying $F_A$ to $\pi\Nc\mA^2$.
This factor is obtained from the calculated
amplitude; it has not been derived independently
from the factorisation formula.

Having fixed this constant, the predicted
dependence on $x$ can be tested away from
threshold. The forward limit $c\to1$ at
fixed velocity also gives $p_T\to0$, but the
produced bosons now move along the beam.
Since the test is also applied to $gg\to HH$ later, we denote the velocity and mass of the produced bosons by $\beta$ and $m$, with $\beta=\bA$ and $m=\mA$ for $AA$.
In this limit, the longitudinal velocity of
the first boson is $2x-1$, so
$x=(1+\beta)/2$.

To extract the logarithmic coefficient, we differentiate the real part
of the one-loop remainder with respect to $\ln(p_T^2/m^2)$ at fixed
$\beta$ and $m$, and then take the forward limit.
Dividing by the normalisation fixed above defines
\begin{align}
 D(\beta)
 &\equiv
 \frac{1}{\pi\Nc m^2}
 \lim_{c\to1}
 \frac{\partial\,\Real\mathcal M_1^{(1)}}
      {\partial\ln p_T^2}
 \nonumber\\
 &=F_A\!\left(\frac{1+\beta}{2}\right)
 \qquad (gg\to AA).
 \label{eq:dps-prediction}
\end{align}
The last equality is the prediction to be
tested: no further adjustment of the
normalisation is made as $\beta$ is varied.
We evaluated $D(\beta)$ from the exact one-loop
remainders using $1-c=10^{-14}$ and
$10^{-18}$, over
$\DPSBetaMin\leq\beta\leq\DPSBetaMax$,
with $\Nc=$ \DPSNcValues{} and
$\nf=$ \DPSNfValues{}.
The largest relative deviation from
Eq.~\eqref{eq:dps-prediction} was
\DPSMaxDeviation{}.

All logarithms in the one-loop remainder
of Ref.~\cite{Bhattacharya:2019bmd} have
positive arguments in the physical region,
and all dilogarithms have arguments between
0 and~1. These functions are therefore real,
and the real part of the remainder follows
from its explicit factors of $i$. It
contains no dilogarithms and simplifies to
\begin{equation}
 \Real\mathcal M_1^{(1)}
 =\pi m^2\left[\Nc\,F_A\!\left(\frac{1+\beta}{2}\right)
 \ln\frac{\beta^2(1-c^2)}{1-\beta^2c^2}
 -\frac{2\beta_0}{1-\beta^2}\right],
 \label{eq:re-one-loop}
\end{equation}
which does not depend on $\muR$.
As a numerical check, it agrees with the
remainder of Ref.~\cite{Bhattacharya:2019bmd} at
\ClosedFormPoints{} points with
$\ClosedFormBetaMin\leq\beta\leq\ClosedFormBetaMax$
and $|c|\leq\ClosedFormCosMax$, for
$(\Nc,\nf)=$ \ClosedFormSets{} and
$(m,\muR)=$ \ClosedFormScales{}~GeV.
With \ClosedFormDigits-digit arithmetic
the relative differences are below
\ClosedFormDev{}. The simplification, in
exact arithmetic, and this check are part
of the released package.
At fixed $\beta$ and $m$,
$\ln p_T^2=\ln(1-c^2)+\text{const}$, so
\[
 \frac{\partial\,\Real\mathcal M_1^{(1)}}
      {\partial\ln p_T^2}
 =\pi\Nc m^2F_A\!\left(\frac{1+\beta}{2}\right)
 \frac{1-\beta^2}{1-\beta^2c^2},
\]
which tends to
$\pi\Nc m^2F_A\bigl((1+\beta)/2\bigr)$
as $c\to1$.
Equation~\eqref{eq:dps-prediction} therefore
holds for all $\beta$. At threshold,
Eq.~\eqref{eq:re-one-loop} gives
$\Real\widehat A^{(1)}_{10}=-\pi(7\Nc\ln2+2\beta_0)$,
in agreement with the value quoted after
Eq.~\eqref{eq:one-loop-threshold}.

\paragraph{Comparison with scalar pair production.}

The scalar process provides a separate test
of the relative signs in the helicity sum.
The splittings and colour factors are
unchanged, while replacing the CP-odd
vertices by CP-even ones changes $F_A$
to $F_S$.
We tested this prediction using the one-loop amplitudes for $gg\to HH$ in which each scalar is emitted from its own effective vertex, at the same mass $m$. We computed them with the code of Ref.~\cite{Bhattacharya:2019bmd}, replacing the CP-odd vertices of $O_G$ by those of the CP-even operator $H\,G^a_{\mu\nu}G^{a\mu\nu}$ and leaving the rest of the calculation unchanged.
The code reproduces the one-loop $AA$
result of that reference to
\HHReproduceDigits{} digits, and the
$HH$ double poles agree with Catani's
formula~\cite{Catani:1998bh}.

In the conventions used here, the $HH$
tree amplitude at threshold is $-2im^2$,
opposite in sign to the $AA$ amplitude.
The corresponding forward coefficient
satisfies
\[
 D(\beta)
 =-F_S\!\left(\frac{1+\beta}{2}\right)
 \qquad (gg\to HH)
\]
to a relative accuracy of \HHDpsMaxDev{}
for $\HHBetaMin\leq\beta\leq\HHBetaMax$.
The normalisation fixed from $AA$ therefore
also reproduces the $HH$ logarithm without
an additional fitted constant. In units
of the respective tree amplitudes, the
ratio of the scalar and pseudoscalar
logarithmic coefficients at threshold is
\begin{equation}
 \dfrac{F_S(\tfrac12)}{F_A(\tfrac12)} = \dfrac{9}{7}.
\end{equation}
The opposite helicity amplitude again
has no logarithm. Its extracted forward
coefficient is about \HHMtwoMax{} times
that of $\mathcal M_1$.
For this check, the diagram with
the four-gluon term of the CP-even
operator and the contact counterterm
for two such insertions were not
included. Neither produces a logarithm
of $p_T$ as $p_T\to0$, so neither changes
the logarithmic coefficient being tested.

\begin{figure}[!htb]
 \centering
 \includegraphics[width=\linewidth,height=0.75\textheight,keepaspectratio]{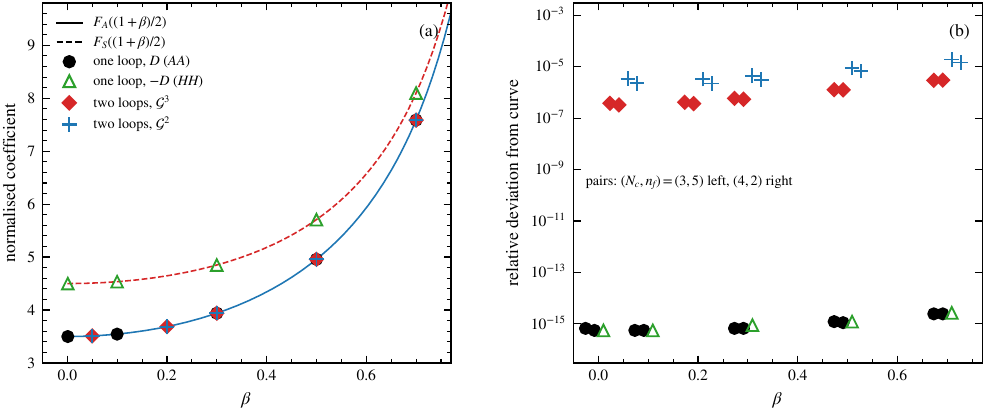}
 \caption{Forward coefficients of the DPS logarithms and the helicity sums $F_A$, $F_S$; each coefficient has the same symbol in both panels. (a) $(\Nc,\nf)=(3,5)$. Circles: $D(\beta)$ for $gg\to AA$, Eq.~\eqref{eq:dps-prediction}. Triangles: $-D(\beta)$ for $gg\to HH$ through two CP-even effective vertices; the sign reversal removes the relative sign of the scalar and pseudoscalar tree amplitudes at threshold. Diamonds and crosses: the real parts of the coefficients of $\mathcal G^3$ and $\mathcal G^2$ in the forward limit of the two-loop remainder of $gg\to AA$, determined from $1-c=10^{-7},\ldots,10^{-10}$ and divided by $-\tfrac43\Nc\cdot2\pi\Nc m^2$ and $-2\beta_0\cdot2\pi\Nc m^2$ (Sec.~\ref{sec:two-loop}). Solid and dashed curves: $F_A((1+\beta)/2)$ and $F_S((1+\beta)/2)$. The values for $(\Nc,\nf)=(4,2)$ coincide with those shown. (b) Relative deviation of each point from its curve in (a), for $(\Nc,\nf)=(3,5)$ and, for $gg\to AA$, $(4,2)$; the symbols are displaced horizontally for legibility.}
 \label{fig:dps}
\end{figure}

Figure~\ref{fig:dps}(a) compares the extracted coefficients with the two helicity sums. The $AA$ points lie on the curve $F_A$ for both $(\Nc,\nf)=(3,5)$ and $(4,2)$, so that after the normalisation of Eq.~\eqref{eq:dps-prediction} the coefficient does not depend on $\Nc$ or $\nf$. The $HH$ points (triangles), plotted with their sign reversed, lie on the curve $F_S$, which is separated from $F_A$ by
\begin{equation}
 F_S-F_A=4x(1-x)=1-\beta^2.
\end{equation}
Panel~(b) shows the relative deviations from the predictions; the two-loop points are discussed in Sec.~\ref{sec:two-loop}.

\paragraph{Colour, scale dependence and phase.}

The factorisation also explains the flavour
and colour dependence of the one-loop
logarithm. The $O_G$ vertices couple only
to gluons, so only gluon splittings enter this
singular configuration and its coefficient is
independent of $\nf$. The factor $\Nc$
comes from the colour contraction of
the two three-gluon vertices.
Renormalisation and IR subtraction
do not change this logarithmic coefficient:
at one loop, their contributions are
proportional to the tree amplitude and
do not generate a logarithm of $p_T$.
The coefficient is therefore independent
of $\muR$.

Finally, the coefficient
$\OneLoopAone\,\mA^2$ of $\mathcal L$ is
real, whereas
$\mathcal M_1^{(0)}\to2i\mA^2$
at threshold. The logarithm is thus
out of phase with the tree amplitude
by $90^\circ$.
This relative phase is consistent
with the absorptive origin of the
one-loop DPS singularity, which can
be obtained from cuts that put
internal lines on
shell~\cite{GauntStirling2011}.
It explains why the one-loop DPS
logarithm does not contribute to
the interference with the tree
amplitude, as discussed in
Sec.~\ref{sec:rate}.

\subsection{Two loops}
\label{sec:two-loop}

The leading logarithms of the calculated two-loop coefficient $\mathcal M_1$ can be written in terms of the one-loop DPS logarithm. Comparing the coefficients of Eq.~\eqref{eq:leading-coefficients} with the coefficient $\OneLoopAone$ of Eq.~\eqref{eq:one-loop-threshold} gives, for general $\Nc$ and $\nf$,
\begin{equation}
 \widehat A_{13}\mathcal L^3+\widehat A_{12}\mathcal L^2
 =\OneLoopAone\,\mathcal L\left[\LogRelationCubic\,\mathcal L^2-2\beta_0\,\mathcal L\right],
 \label{eq:log-relation}
\end{equation}
with $\beta_0$ as in Eq.~\eqref{eq:normalisation}.
A similar relation holds in planar $\mathcal N=4$ super Yang--Mills theory.
The six-gluon maximally helicity violating (MHV) amplitude in planar \(\mathcal N=4\) super Yang--Mills theory, with two negative and four positive helicities when all momenta are outgoing, was studied in Refs.~\cite{DixonEsterlis2016,DixonEsterlis2016Erratum}.
Near the configuration that mimics DPS, its singular terms obey an evolution equation that fixes the leading logarithms to all orders~\cite{DixonEsterlis2016,DixonEsterlis2016Erratum}. Its derivation follows the analysis of crossing Wilson lines in Ref.~\cite{KorchemskayaKorchemsky1995}.
For this comparison we use the nonsingularly framed hexagonal
Wilson loop $\mathcal W^{\mathrm{ns}}_{3\to3}$ of
Ref.~\cite{DixonEsterlis2016}. Its leading-logarithmic series,
Eq.~(6.2) of that reference, gives the ratio of the leading
two-loop and one-loop terms, including their powers of the
coupling, as $-(a/6)\ln^2|\delta|$.
Equation~(2.37) of the same reference relates this loop to the
BDS-like normalised MHV amplitude through a factor that is
independent of $\delta$ at fixed nonsingular kinematics and
starts at unity in the coupling expansion. It therefore leaves
the highest logarithmic terms at one and two loops unchanged.
Here $a=g^2\Nc/(8\pi^2)$ and $|\delta|$, which measures the
approach to the singular configuration, is proportional to
the squared transverse momentum at fixed values of the other
kinematic variables. With $a=2\Nc\as$ and
$\ln|\delta|=2\mathcal L+\mathrm{const}$, the ratio becomes
$-\tfrac43\Nc\as\mathcal L^2$ at leading logarithmic order.
This is the ratio of the cubic term in
Eq.~\eqref{eq:log-relation} to the one-loop logarithm, once the
additional power of $\as$ at two loops is included.
The second term, $-2\beta_0\mathcal L$, has no counterpart in the
$\mathcal N=4$ theory; the number of quark
flavours enters it only through $\beta_0$.

A corresponding $\beta_0$ coefficient appears in the NLO momentum-space
kernel for $g\to gg$ splitting in unpolarised colour singlet double
parton distributions~\cite{Diehl:2019rdh}.
There the logarithm is $L_\Delta=\ln(\mu^2/\Delta^2)$, where $\Delta$ is
the transverse momentum mismatch between amplitude and conjugate
amplitude, rather than the transverse momentum of a produced particle.
The LO kernel contains the LO splitting function multiplied by
$L_\Delta$, and Eqs.~(50) and (135) of Ref.~\cite{Diehl:2019rdh} give a
$\beta_0$ contribution to the coefficient of $L_\Delta^2$ equal to
$\beta_0/2$ times the same splitting function, in units of
$\alpha_s/2\pi$. Relative to the LO logarithmic term this is
$(\alpha_s/2\pi)(\beta_0/2)L_\Delta=\as\beta_0L_\Delta$, which the
replacement $L_\Delta\to-2\mathcal L$ turns into the term
$-2\beta_0\mathcal L$ of Eq.~\eqref{eq:log-relation}, with the
additional power of $\as$ included.
Half of this coefficient comes from the renormalisation of the coupling
and half from the $\delta$-function parts of the DGLAP kernels, so that
the running of the coupling alone would give $-\beta_0\mathcal L$.
This is a comparison of coefficients, not an identification of $\Delta$
with $p_T$. The other terms of the kernel are not part of it: a term
proportional to $C_A\ln[x(1-x)]$ from the plus distributions, which
contributes where the two momentum fractions add up to one, and the
real emission terms, which contribute where they add up to less than one.

The following tests indicate that the relation of Eq.~\eqref{eq:log-relation} extends beyond threshold. In the forward limit at fixed velocity a convenient variable is $\mathcal G=\tfrac12\ln[(\mA^4-tu)/\mA^4]=\mathcal L-\ln(1-\bA^2)$, in which the one-loop DPS logarithm is $C_1\mathcal G$ and the relation reads
\begin{align}
 \mathcal M_1^{(2)}&\supset C_1\Bigl[-\frac43\Nc\,\mathcal G^3-2\beta_0\,\mathcal G^2\Bigr],
 \label{eq:forward-relation}
 \\
 C_1&=2\pi\Nc \mA^2F_A\Bigl(\frac{1+\bA}{2}\Bigr),
 \nonumber
\end{align}
with $C_1$ from Eq.~\eqref{eq:dps-prediction}. In the threshold series, Eq.~\eqref{eq:forward-relation} holds exactly through $\bA^6$ for general $\Nc$ and $\nf$; at fixed angle, the coefficient of $\mathcal L^3$ follows the expansion of $F_A\bigl((1+\bA)/2\bigr)$ order by order.

Away from threshold we evaluated the released two-loop remainder of $\mathcal M_1$ at $1-c=10^{-6},\ldots,10^{-10}$ with \MoneFwdDigits-digit arithmetic. For each velocity and colour/flavour set, the point at $1-c=10^{-10}$ was reevaluated with \MoneFwdCheckDigits-digit arithmetic; these comparisons agree to at least \MoneFwdCheckAgreement{} digits. The four points with $1-c\leq10^{-7}$ determine the four coefficients of a cubic polynomial in $\ln(1-c)$. For $\MoneFwdBetaMin\leq\beta\leq\MoneFwdBetaMax$ and $(\Nc,\nf)=$ \MoneFwdSets, the real parts of the resulting coefficients of $\mathcal G^3$ and $\mathcal G^2$ agree with Eq.~\eqref{eq:forward-relation} to \MoneFwdDevThree{} and \MoneFwdDevTwo{} (Fig.~\ref{fig:dps}), and their imaginary parts are below \MoneFwdImag{} of their magnitudes. A least-squares fit that also includes the point at $1-c=10^{-6}$ gives deviations about \MoneFwdWindowFactor{} times larger, as expected from power corrections in $1-c$. Within the range tested, the two-loop leading logarithms of $\mathcal M_1$ in the forward limit are therefore the one-loop DPS logarithm, with its helicity sum $F_A(x)$, multiplied by the same factor at every velocity.

The $|J_z|=2$ amplitude acquires a logarithm at two loops: the term
$\widehat A_{22}\mathcal L^2$, with
$\widehat A_{22}=(\MtwoRelation)\cdot\OneLoopMtwo$, is the one-loop
constant multiplied by $\MtwoRelation\,\mathcal L^2$. The forward-limit calculation described next indicates that the same coefficient relation persists away from threshold. In the forward limit $c\to1$ at fixed velocity, where $p_T\to0$ as well, we evaluated the released two-loop remainder of $\mathcal M_2$ at $1-c=$ \JzExponents{} with \JzDigits-digit arithmetic. For each velocity and colour/flavour set, the most forward fitted point, $1-c=10^{-7}$, was reevaluated with \JzCheckDigits-digit arithmetic; these comparisons agree to at least \JzCheckAgreement{} digits. These three points determine the three coefficients of a quadratic polynomial in $\ln(1-c)$. For $\JzBetaMin\leq\beta\leq\JzBetaMax$ and $(\Nc,\nf)=$ \JzSets, the real part of the coefficient of $\ln^2(2p_T/m)$ equals $-4\Nc$ times the forward limit of $\mathcal M_2^{(1)}$ to \JzMaxDev, although that limit varies by a factor of \JzKOneRatio{} over this range. Its imaginary part, which vanishes in the proposed coefficient relation, is below \JzImag{} of its magnitude. A least-squares fit that also includes the point at $1-c=10^{-4}$ increases the deviation to \JzMaxDevWide, as expected from power corrections. Within the range tested, the one-loop amplitude and the extracted two-loop logarithm are therefore consistent with
\begin{equation}
 \as\mathcal M_2^{(1)}\bigl[1-4\Nc\as\ln^2(2p_T/m)\bigr],
 \label{eq:jz2-forward}
\end{equation}
with no additional velocity or flavour dependence resolved in the coefficient of the double logarithm.

The first-order expansion of $\exp[-4N_ca_s\ln^2(2p_T/m)]$, multiplied by the one-loop amplitude, reproduces Eq.~\eqref{eq:jz2-forward} at double-logarithmic accuracy.
At threshold, integrating $-4\Nc\mathcal L^2$ with respect to $\mathcal L$ and multiplying by $\OneLoopAone$ reproduces the cubic term of Eq.~\eqref{eq:log-relation}.
These comparisons concern the highest logarithmic powers only.
The exponent also agrees at leading order with $-\gamma_K\ln^2\delta/8$ in Ref.~\cite{DixonEsterlis2016,DixonEsterlis2016Erratum}, using $\gamma_K=4a$, $a=2N_ca_s$ and $\ln\delta=2\mathcal{L}+\mathrm{const}$ at leading-logarithmic accuracy.
These coefficient relations do not establish exponentiation in QCD.
%
%
%
%
The $J_z$ argument of
Sec.~\ref{sec:dps} concerns the one-loop singularity and does not forbid
this logarithm. Nor can it be explained by loop corrections to the $gg\to A$ subprocess amplitudes, since conservation of angular momentum forbids $gg\to A$ for opposite gluon helicities to all orders; its dynamical origin remains open. A related
effect occurs in planar $\mathcal N=4$ super Yang--Mills theory.
The next-to-MHV (NMHV) six-gluon amplitude, with three negative helicities, has no such singularity at one loop because of \(J_z\) conservation. Logarithmic singularities starting at two loops were found in the region where this limit overlaps with multi-Regge kinematics. Contributions in which more than one particle crosses a cut were suggested as their source~\cite{DixonEsterlis2016,DixonEsterlis2016Erratum}.

In the forward limit the one-loop remainders agree with
\begin{equation}
 \lim_{c\to1}\frac{\mathcal M_2^{(1)}}{m^2}=\pi\Nc\,W(x),\qquad
 W(x)=\frac{x^4+(1-x)^4}{x(1-x)},\qquad x=\frac{1+\beta}{2},
 \label{eq:K1}
\end{equation}
to \KOneClosedDev{} for $\KOneBetaMin\leq\beta\leq\KOneBetaMax$ and both sets of $(\Nc,\nf)$; at threshold it gives $\OneLoopMtwo$.
With the factors of Eq.~\eqref{eq:splitting-weights},
$W(x)=w_{+-}(x)w_{-+}(1-x)+w_{-+}(x)w_{+-}(1-x)$.
This is the sum over the two helicity pairings allowed for an
opposite-helicity initial state, with equal gluon helicities at each
effective vertex (Sec.~\ref{sec:dps}).
The prefactor $\pi\Nc$ in Eq.~\eqref{eq:K1} also matches that in
Eq.~\eqref{eq:dps-prediction}.
This structure suggests a DPS origin for the finite one-loop forward
limit of $\mathcal M_2$, with the azimuthal phase removing the
logarithm but leaving a finite term.
We have not derived this term from the factorisation formula.
The two helicity amplitudes provide an explicit two-loop QCD
benchmark for a general description of this singular region.

Figure~\ref{fig:amplitudes} compares the three orders. With $\as=\BenchAs$ ($\alpha_s=\BenchAlphaS$), at $\bA=\LoopSizeBeta$ and $c=\BenchCos$, the one- and two-loop terms of $\mathcal M_1$ are \MoneOneLoopRatio{} and \MoneTwoLoopRatio{} times the tree term, and the one-loop term of $\mathcal M_2$ is \MtwoOneLoopRatio{} times the tree term. The last ratio grows as $\bA^{-2}$ towards threshold, since the tree-level $|J_z|=2$ amplitude vanishes there. The one-loop terms include the scale term of Eq.~\eqref{eq:normalisation}. Towards threshold, the loop terms of $\mathcal M_1$ grow with $|\mathcal L|$, the two-loop term as $|\mathcal L|^3$ and the one-loop term as $|\mathcal L|$, so that below $\bA\approx\CrossoverBeta$ the two-loop term exceeds the one-loop term at this angle. Because the logarithms are out of phase with the tree amplitude, the strongest of them are absent from the finite virtual correction (Sec.~\ref{sec:rate}). In panel (b) the one-loop term of $\mathcal M_2$ tends to the constant of Eq.~\eqref{eq:one-loop-threshold}, while the two-loop term grows as $\mathcal L^2$.

%% file: tables/leading_logs.tex
\begin{table*}[!htb]
\centering

\begin{tabular}{cclrr}
\toprule
 & $k$ & exact & $\operatorname{Re}\widehat A_{ik}$ & $\operatorname{Im}\widehat A_{ik}$ \\
\midrule
$\mathcal M_1$ & $3$ & $-\tfrac{28}{3}\pi N_c^2$ & $-2.63894\times10^{2}$ & $0$ \\
$\mathcal M_1$ & $2$ & $-\tfrac{14}{3}\pi N_c(11N_c-2n_f)$ & $-1.01159\times10^{3}$ & $0$ \\
$\mathcal M_1$ & $1$ & Eq.~\eqref{eq:M1-L1} & $3.55129\times10^{3}$ & $4.76702\times10^{3}$ \\
$\mathcal M_1$ & $0$ & -- & $-3.98215\times10^{2}$ & $-4.00569\times10^{3}$ \\
$\mathcal M_2$ & $2$ & $-2\pi N_c^2$ & $-5.65487\times10^{1}$ & $0$ \\
$\mathcal M_2$ & $1$ & $-\tfrac{4}{3}\pi N_c(4N_c-n_f)$ & $-8.79646\times10^{1}$ & $0$ \\
$\mathcal M_2$ & $0$ & -- & $3.12007\times10^{2}$ & $3.40501\times10^{2}$ \\
\bottomrule
\end{tabular}
\caption{Coefficients of the leading power term, $\mathcal M_i|_{\beta_A^0}=m_A^2\sum_k \widehat A_{ik}\,\mathcal L^k$, with $\mathcal L$ of Eq.~\eqref{eq:calL}. They do not depend on the scattering angle. Exact forms for $\mu_R=m_A$ and general $N_c$, $n_f$; numerical values for $N_c=\BenchNc$, $n_f=\BenchNf$, $\mu_R=m_A$.}
\label{tab:leading-log-benchmark}
\end{table*}

%% file: sections/05_rate.tex
\section{Consequences for the cross section}
\label{sec:rate}

\subsection{Leading order}

\begin{figure*}[!htb]
 \centering
 \includegraphics[width=0.94\textwidth]{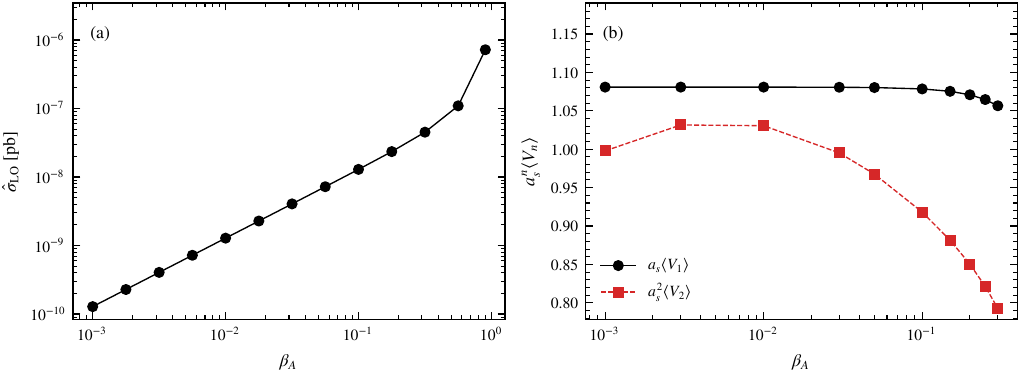}
 \caption{(a) Leading order partonic cross section of the Type~II amplitude, Eq.~\eqref{eq:dsigma}, for $\mA=\BenchMA\,\mathrm{GeV}$, $\alpha_s=\BenchAlphaS$ and $\cot\beta=\BenchCotBeta$. (b) Tree-weighted angle averages $a_s\langle V_1\rangle$ and $a_s^2\langle V_2\rangle$ of the finite virtual corrections, Eq.~\eqref{eq:angle-average}, for $\mu_R=m_A$, $\Nc=\BenchNc$ and $\nf=\BenchNf$. Real emission is not included in panel (b).}
 \label{fig:cross-section}
\end{figure*}

Averaging Eq.~\eqref{eq:helicity} over helicities and colours, and including a factor $\tfrac12$ for identical particles, gives the LO partonic cross section of the Type~II amplitude,
\begin{equation}
 \frac{\dd\hat\sigma}{\dd c}
 =\frac{\bA\,C_G^4}{128\pi\hat s\,(\Nc^2-1)}\left(|\mathcal M_1^{(0)}|^2+|\mathcal M_2^{(0)}|^2\right)
 \label{eq:dsigma}
\end{equation}
for $-1<c<1$. For $\alpha_s=\BenchAlphaS$, $\cot\beta=\BenchCotBeta$ and $\mA=\BenchMA\,\mathrm{GeV}$, we find $\hat\sigma=\SigmaLO\,\mathrm{pb}$ at $\bA=\SigmaBeta$ (Fig.~\ref{fig:cross-section}(a)). The cross section rises as $\bA$ above threshold and as $\hat s$ at large $\hat s$, as expected for dimension-five operators, and it is of fourth order in $\as$ through $C_G^4$. In a 2HDM, it is a small part of the total, which is dominated by Type~I (Sec.~\ref{sec:counting}).

\subsection{Finite virtual corrections}

To examine the effect of the loop corrections on the Type~II squared amplitude, we define two dimensionless
coefficients, $V_1$ and $V_2$, relative to the tree contribution. Let $\mathcal A_{\Fin}$ denote the
amplitude constructed from the finite remainders of Eq.~\eqref{eq:catani}, in the normalisation of
Eq.~\eqref{eq:normalisation}. Using the helicity decomposition in Eq.~\eqref{eq:helicity}, its square
can be written as
\begin{align}
 \sum_{\mathrm{hel,col}}|\mathcal A_{\Fin}|^2
 &=2(\Nc^2-1)\,C_G^4
   \sum_i|\mathcal M_i^{(0)}|^2
 \times
 \left[1+\as V_1+\as^2V_2+\Ord(\as^3)\right], \text{ where }
 \nonumber\\
 V_1
 &=\frac{
  2\Real\sum_i
  \mathcal M_i^{(0)*}\mathcal M_i^{(1)}
 }{
  \sum_i|\mathcal M_i^{(0)}|^2
 },
 \nonumber\\
 V_2
 &=\frac{
  \sum_i\left(
   2\Real\bigl[\mathcal M_i^{(0)*}\Mfull{i}\bigr]
   +|\mathcal M_i^{(1)}|^2
  \right)
 }{
  \sum_i|\mathcal M_i^{(0)}|^2
 }.
 \label{eq:virtual}
\end{align}
The sums run over the two tensor coefficients, $i=1,2$. The coefficient $V_1$ contains the
interference between the tree and one-loop amplitudes. At the next order, $V_2$ contains
both the interference between the tree and two-loop amplitudes and the square of the
one-loop amplitude.
These coefficients refer only to the $O_G$--$O_G$ Type~II contribution; they are the NLO and NNLO finite virtual terms when the Type~II tree amplitude is the leading amplitude (Sec.~\ref{sec:counting}).
The real emission contributions are not included here. The quantities
$V_1$ and $V_2$ therefore retain the dependence on the subtraction scheme
of Eq.~\eqref{eq:catani}, which cancels only when the real emission and the
counterterms for initial state collinear factorisation are added; they are
not observables or $K$ factors. The coefficient of the squared logarithm derived below is not affected by the finite changes of the subtraction scheme specified below.

At fixed $\bA$, both coefficients depend on the scattering angle. To display
their velocity dependence in Fig.~\ref{fig:cross-section}(b), we average them
over the angle with the Type~II LO cross section as the weight. Since the
remaining factors in Eq.~\eqref{eq:dsigma} depend only on $\bA$ and cancel
in this average, $\as\langle V_1\rangle$ and $\as^2\langle V_2\rangle$ are
the finite virtual contributions relative to the angle-integrated tree
contribution, rather than unweighted averages. The weighted average is
\begin{equation}
 \langle V_n\rangle
 =
 \frac{
  \displaystyle
  \int_{-1}^{1}\dd c\,
  \sum_i|\mathcal M_i^{(0)}|^2\,V_n
 }{
  \displaystyle
  \int_{-1}^{1}\dd c\,
  \sum_i|\mathcal M_i^{(0)}|^2
 },
 \qquad n=1,2.
 \label{eq:angle-average}
\end{equation}

The angular integrals are evaluated with \QuadNodes-point Gauss--Legendre quadrature; with \QuadNodesRef{} points, $\langle V_1\rangle$ and $\langle V_2\rangle$ change by relative amounts smaller than \QuadVoneDev{} and \QuadVtwoDev{} at $\bA=$ \QuadBetas.
We evaluate the one-loop remainders without a threshold expansion and the
two-loop remainders with the series through $\bA^6$. For
$\alpha_s=\BenchAlphaS$ and $\muR=\mA=\BenchMA\,\mathrm{GeV}$, the curves of
Fig.~\ref{fig:cross-section}(b) lie between \VoneMin{} and \VoneMax{} for
$\as\langle V_1\rangle$, and between \VtwoMin{} and \VtwoMax{} for
$\as^2\langle V_2\rangle$, over $\VirtualBetaMin\leq\bA\leq\VirtualBetaMax$.
At fixed kinematics, the scale dependence of the one-loop remainder in
Eq.~\eqref{eq:normalisation} gives $\partial V_1/\partial\ln\muR^2=4\beta_0$.

\paragraph{One-loop contribution.}

The threshold behaviour of $V_1$ and $V_2$ follows from the relative phases
of the tree and loop amplitudes. At fixed scattering angle,
$\mathcal M_1^{(0)}\to2i\mA^2$, whereas $\mathcal M_2^{(0)}$ vanishes as
$\bA^2$. The squared tree amplitude in the denominator of
Eq.~\eqref{eq:virtual} therefore tends to $4\mA^4$, and at leading power
interference with the tree amplitude selects the imaginary part of
$\mathcal M_1^{(1)}$ or $\Mfull{1}$. With $\ell=\ln(2p_T/\mA)$,
Eq.~\eqref{eq:calL} gives $\mathcal L=\ell+i\pi/2+\Ord(\bA^2)$, and the
one-loop result of Eq.~\eqref{eq:one-loop-threshold} reads
\[
 \frac{\mathcal M_1^{(1)}}{\mA^2}
 =
 \OneLoopAone\,\ell
 +\frac{i\pi}{2}\OneLoopAone
 +\widehat A^{(1)}_{10}.
\]
The coefficient of $\ell$ is real, so the logarithm drops out of the
interference with the imaginary tree amplitude. Only the imaginary
constants remain:
\begin{equation}
 V_1\longrightarrow
 \frac{\pi}{2}\OneLoopAone
 +\Imag\widehat A^{(1)}_{10}.
 \label{eq:v1-threshold}
\end{equation}
For $\Nc=\BenchNc$, $\nf=\BenchNf$ and $\muR=\mA$, this gives
$V_1\to\VoneThreshold$. The one-loop remainder thus contains the DPS
logarithm, but the leading power finite virtual correction does not. The
limit is also independent of the angle, which is why the angular average in
Fig.~\ref{fig:cross-section}(b) is nearly constant.

\paragraph{Two-loop contribution.}

The interference of the two-loop amplitude with the tree amplitude again
selects the imaginary part of the loop remainder, but the square of the
one-loop amplitude retains the real logarithm. Using
Eqs.~\eqref{eq:leading-power-M1}, \eqref{eq:leading-power} and \eqref{eq:one-loop-threshold}, the leading
power terms of Eq.~\eqref{eq:virtual} are
\begin{align}
 V_2\big|_{\bA^0}
 &=
 \Imag\left[
  \sum_{k=0}^{3}\widehat A_{1k}\mathcal L^k
 \right] +
 \frac14\left|
  \OneLoopAone\,\mathcal L
  +\widehat A^{(1)}_{10}
 \right|^2
 +\frac{\pi^2\Nc^2}{16}.
\end{align}
The three terms are the interference of the two-loop amplitude with the
tree amplitude, the square of $\mathcal M_1^{(1)}$, and the square of
$\mathcal M_2^{(1)}\to\pi\Nc\mA^2/2$. The last contributes at leading power
although the corresponding tree amplitude vanishes, while the interference
of $\mathcal M_2^{(0)}$ with $\Mfull{2}$ is suppressed by $\bA^2$, up to
logarithms.

The coefficients $\widehat A_{13}$ and $\widehat A_{12}$ are real, and at
leading power
\[
 \Imag\mathcal L^3
 =\frac{3\pi}{2}\ell^2-\frac{\pi^3}{8},
 \qquad
 \Imag\mathcal L^2=\pi\ell.
\]
The cubic logarithm of the two-loop amplitude therefore gives only a
quadratic logarithm in its interference with the tree amplitude, and
$\widehat A_{12}\mathcal L^2$ only a linear one. The square of the one-loop
amplitude is at most quadratic in $\ell$. The leading threshold expansion
of $V_2$ therefore has the form
\begin{equation}
 V_2=v_0+v_1\ell+v_2\ell^2
     +\Ord_{\log}(\bA^2),
 \label{eq:v2-form}
\end{equation}
where the subscripts of $v_0$, $v_1$ and $v_2$ give the power of $\ell$,
not the number of loops.

The coefficient $v_2$ receives two contributions: $\tfrac32\pi\widehat
A_{13}=-14\pi^2\Nc^2$ from the interference of the two-loop amplitude with
the tree amplitude, and $\tfrac14(\OneLoopAone)^2=\tfrac{49}{4}\pi^2\Nc^2$
from the square of the one-loop amplitude. Their sum is
\begin{equation}
 v_2
 =-14\pi^2\Nc^2+\frac{49}{4}\pi^2\Nc^2
 =-\frac74\pi^2\Nc^2.
 \label{eq:v2-log}
\end{equation}
The two contributions partly cancel. At leading power, the DPS logarithm
therefore first appears in the finite virtual correction in $V_2$, with
highest power $\ell^2$.
A fit of the numerical values of $V_2$ at $c=\BenchCos$ over
$\VtwoBetaSmall\leq\bA\leq\VtwoBetaLarge$ gives $v_2=\VtwoLogFit$, while
Eq.~\eqref{eq:v2-log} gives $\VtwoLogExact$ for $\Nc=\BenchNc$; the
relative difference before rounding is \VtwoLogRelDiff{}. For
$\as=\BenchAs$, the quadratic term in the finite virtual correction is
$\as^2v_2\ell^2=\VtwoLogScaled\,\ell^2$.

\paragraph{Remaining coefficients and numerical behaviour.}

At leading power, $v_0$, $v_1$ and $v_2$ are independent of the scattering
angle. The angular dependence of $V_2$ itself remains through $\ell$.
Fits for $v_0$ and $v_1$ at $c=0$, $\tfrac13$ and $\tfrac23$
agree within a relative spread of \VtwoAngleSpread{}.
The coefficient of the linear logarithm follows from the same expansion:
\begin{align*}
 v_1
 &=\pi\widehat A_{12}
   +\Imag\widehat A_{11}
   +\frac{\OneLoopAone}{2}
    \Real\widehat A^{(1)}_{10}
 \\
 &=\frac72\pi\Nc\Real\widehat A^{(1)}_{10}
   +7\pi^2\Nc\beta_0.
\end{align*}
With $\Real\widehat A^{(1)}_{10}=-\pi(7\Nc\ln2+2\beta_0)$ from
Eq.~\eqref{eq:re-one-loop}, the terms proportional to $\beta_0$ cancel and
$v_1=-\tfrac{49}{2}\pi^2\Nc^2\ln2$.
For $\Nc=\BenchNc$ this gives $v_1=\VoneFit$, in agreement with
the numerical fit to a relative difference of \VoneFormulaDev{}.
Neither $v_1$ nor $v_2$ depends on $\nf$ or $\muR$. At leading power, the explicit scale dependence of $V_2$
is therefore confined to $v_0$, which changes from \VzeroBench{}
to \VzeroTwoMu{} when $\muR$ is increased from $\mA$ to $2\mA$.

Although $\ell^2$ is the highest power of the logarithm, it need
not give the largest contribution at the velocities considered.
At $c=\BenchCos$, the ratio $(v_1\ell+v_2\ell^2)/V_2$ is
approximately \LogShareLarge\% at $\bA=\LogShareBetaLarge$
and \LogShareSmall\% at $\bA=\LogShareBetaSmall$.
The quadratic term alone contributes $-\SquareShareLarge\%$ and
$-\SquareShareSmall\%$ of the total $V_2$, respectively.
Since $\ell<0$ in this region and both $v_1$ and $v_2$ are
negative, the linear term is positive and the quadratic term is negative.
This competition also explains the shape of the correction as a
function of velocity. The leading power expression has a maximum at
\[
 \ell_{\mathrm{max}}
 =-\frac{v_1}{2v_2}
 =-7\ln2
 =\VtwoPeakEll,
\]
that is, $p_T=\mA/256$.
At $c=\BenchCos$, this corresponds to $\bA\simeq\VtwoPeakBeta$.
For larger $\bA$, $\ell$ increases and the leading power value of
$V_2$ decreases. The same balance between the linear and quadratic
terms explains the turnover of the angular average
$\as^2\langle V_2\rangle$ in Fig.~\ref{fig:cross-section}(b).
At $c=\BenchCos$ and $\bA=\LogShareBetaLarge$,
including the terms beyond leading power changes $V_2$ by only \LeadPowerDevLarge\%.

\paragraph{DPS interpretation and subtraction dependence.}

The two contributions to $v_2$ also differ in their relation to the DPS
configuration. In the square of the one-loop amplitude, the DPS
region occurs in both the amplitude and its complex conjugate, that is,
on both sides of the final state cut. This is the configuration
associated with perturbative splittings in both protons that
overlaps with DPS and requires a subtraction to avoid double
counting~\cite{DiehlGauntSchoenwald2017,DiehlPloessl2026}.
In the interference with the tree amplitude, the DPS region occurs
only on the loop side of the cut. Both contributions enter the
finite virtual coefficient $V_2$ defined in Eq.~\eqref{eq:virtual}; the
logarithmic term in its one-loop square alone would therefore
not give the complete coefficient $v_2$.

Although $V_2$ depends on the IR subtraction convention, $v_2$ is unchanged by finite
scheme changes that multiply lower order amplitudes by angle independent functions
regular at the pair threshold. The tree amplitude has no
logarithm, and the one-loop remainder contains at most one power of $\ell$. Such changes can therefore
affect the constant and linear terms in $V_2$, but not the coefficient of $\ell^2$.

Finally, with the Wilson coefficient counting specified in
Sec.~\ref{sec:counting}, the one-loop $O_G$--$O_J$ amplitude contributes
at the same order as the two-loop $O_G$--$O_G$ amplitude.
Its coefficient $C_J$ depends on $m_t$ in the heavy top theory~\cite{Bhattacharya:2019bmd}.
The mixed amplitude is proportional to the tree amplitude by Eq.~\eqref{eq:gj-proportional}.
After including its Wilson coefficient and matching the normalisation, its interference with the tree amplitude gives a kinematically constant shift to $V_2$.
It changes $v_0$, but not $v_1$ or $v_2$, and is not included in the results above.

\subsection{Use in higher order calculations}
\label{sec:use}

For the two-body final state, $m_{AA}=\sqrt{\hat s}=2\mA/\sqrt{1-\bA^2}$.
The range $\bA\leq\RecommendedBeta$ therefore covers pair invariant masses
up to $\ThresholdPercentPointTwo\%$ above threshold; for
$\mA=\BenchMA\,\mathrm{GeV}$, this is $m_{AA}\leq\SqrtSPointTwo\,\mathrm{GeV}$.
The point $\bA=\EdgeBeta$ lies $\ThresholdPercentPointThree\%$ above threshold.
We recommend the series through $\beta_A^6$ for $0<\beta_A\leq0.2$
at the angles tested in Tables~\ref{tab:validation-grid} and \ref{tab:forward-angles}.
At $\bA=\EdgeBeta$, the largest relative error of the complete coefficients on the validation grid is \PtwoErrAtPointThree{}.
These checks do not establish a uniform accuracy bound as $c\to\pm1$, where
the logarithms diverge. The series retain the complex values of both tensor
coefficients, so they can be used in interference with other amplitudes,
provided these are combined with the same normalisation, Wilson coefficient
conventions, tensor basis and infrared subtraction scheme.

One application is the hard function of a threshold resummed or
soft-virtual prediction for the pair invariant mass distribution. There,
threshold refers to the limit of little energy available for additional
radiation at fixed $m_{AA}$, not to
$\bA\to0$~\cite{Sterman:1986aj,Catani:1989ne,deFlorian:2015moa}. At leading power in the soft limit, the soft and collinear factors are universal for gluon fusion into a colour singlet,
and the process dependence enters through the finite virtual contributions
at each $m_{AA}$~\cite{deFlorian:2012za,Bhattacharya:2021nsv}; the series
supply the two-loop Type~II input near $m_{AA}=2\mA$. When the Type~II tree
amplitude is the leading amplitude, the required combinations are those of
Eq.~\eqref{eq:virtual}, and integrated over the angle at fixed $m_{AA}$
they give the averages of Eq.~\eqref{eq:angle-average};
Ref.~\cite{deFlorian:2015moa} gives the corresponding construction for
Higgs pairs.

In a 2HDM, these remainders first contribute at N$^3$LO through
interference with the leading Type~I amplitude
(Sec.~\ref{sec:counting}).
This interference must be formed before angular integration, so the
averages in Eq.~\eqref{eq:angle-average} cannot be used directly.
All remaining amplitudes at this order are also required, including
any mixed $O_G$--$O_J$ contribution.
For Higgs pairs, the NNLO virtual corrections likewise involve
diagrams with two effective vertices only at tree level and one
loop~\cite{deFlorianMazzitelli2013Virtual}.
Their two-loop terms are an ingredient of
N$^3$LO~\cite{Banerjee:2018lfq}.

%% file: sections/06_conclusions.tex
\section{Conclusions}
\label{sec:conclusions}

We have expanded the two-loop Type~II finite remainders for $gg\to AA$
with two $O_G$ insertions about the pair production threshold through
$\bA^6$ at fixed scattering angle.
After reducing the endpoint constants to a basis of \BasisCountWord{}
transcendental constants with numerically identified relations, all
inverse powers of $\bA$ and odd powers
through $\bA^5$ cancel identically. Every coefficient is even in
$\cos\theta$.
At the tested points with $0<\bA\leq\RecommendedBeta$, including points
close to the beam axis ($|c|\leq\ForwardCosMax$), the series reproduce the
two complete remainders with maximum
relative errors of \PoneErrLePointTwo{} and \PtwoErrLePointTwo{},
respectively.
The series files are a factor of \SizeRatio{} smaller than those of
Ref.~\cite{Bhattacharya:2019bmd}.
After conversion using Eq.~\eqref{eq:normalisation}, the remainders pass
the renormalisation group checks of Eqs.~\eqref{eq:rg-one} and
\eqref{eq:rg-two} (App.~\ref{app:normalization}).

At leading power the angular dependence enters only through the logarithm
of the transverse momentum, which we identified with the DPS singularity of
Ref.~\cite{GauntStirling2011}. The $g\to gg$ wavefunction factors and the
relative sign of the CP-odd subprocess amplitudes fix the momentum fraction
dependence of its coefficient; with the overall constant fixed at
threshold, this reproduces the one-loop coefficient in the forward limit
exactly, as Eq.~\eqref{eq:re-one-loop} shows, and a one-loop $gg\to HH$
calculation confirms
the CP-even helicity sum with the same constant.
At two loops the $J_z=0$ and $|J_z|=2$ remainders develop leading logarithms $\mathcal{L}^3$ and $\mathcal{L}^2$, respectively.
The equal-helicity leading-logarithmic ratio agrees with the corresponding result in planar $\mathcal{N}=4$ super Yang--Mills theory \cite{DixonEsterlis2016,DixonEsterlis2016Erratum}, and the comparison with the NLO $g\to gg$ splitting kernel for double parton distributions concerns its $\beta_0$ coefficient \cite{Diehl:2019rdh}.
In the forward variable $\mathcal{G}$, fits support Eq.~\eqref{eq:forward-relation} at the tested velocities in $0.05\leq\beta_A\leq0.7$.
For the opposite-helicity remainder, fits at the tested velocities in $0.05\leq\beta_A\leq0.5$ support a real double-logarithmic coefficient equal to $-4N_c$ times the one-loop forward remainder of Eq.~\eqref{eq:K1}.
At leading power, the one-loop DPS logarithm drops out of the tree interference because of their relative phase.
At the next order, the cubic logarithm of the two-loop remainder gives a quadratic logarithm in this interference.
Combining the interference with the one-loop square gives $V_2\supset-\frac{7}{4}\pi^2N_c^2\ln^2(2p_T/m_A)$.

Several questions remain open: the derivation of the overall factor
$\pi\Nc m^2$ from the factorisation formula with the effective vertices,
the mechanism that produces the $|J_z|=2$ double logarithm, and whether the
observed factor exponentiates. A general treatment of this singular region
would have to reproduce the different logarithmic powers of the two
amplitudes, their colour and flavour dependence and the reduced singularity
of the finite virtual correction; the present results provide explicit
two-loop QCD benchmarks for it, and the known two-loop $gg\to HH$
amplitudes with two effective vertices~\cite{Banerjee:2018lfq} permit the
same comparison for the scalar case.

The series provide the two-loop Type~II input to the hard function for the
pair invariant mass distribution near $m_{AA}=2\mA$, at NNLO for a
pseudoscalar coupled to gluons only through $G\tilde G$ and at N$^3$LO in a
2HDM (Sec.~\ref{sec:use}).

%% file: sections/data_availability.tex
\section*{Data Availability}
\label{sec:dataavailability}

The threshold series, construction files, construction and validation code,
outputs of the original runs, validation records, code for numerical evaluation, endpoint constant reductions, tree-level projections, the one-loop $gg\to HH$ coefficients used in Sec.~\ref{sec:dps} and the scripts that reproduce all tables and figures form one package, in which the one-loop $gg\to HH$ coefficients enter as computed values.
The updated version is provided as ancillary files with the arXiv version of this article and as Supplemental Material~\cite{SupplementalMaterial}.
It is maintained at \url{https://github.com/abhattacharya1992/ggAA-threshold} and archived at Zenodo, \href{https://doi.org/10.5281/zenodo.23597489}{doi:10.5281/zenodo.23597489}.
The unexpanded finite remainders of Ref.~\cite{Bhattacharya:2019bmd} are downloaded from arXiv by the package and checked against their checksums.
The released script \texttt{scripts/check\_dps\_hh.py}
compares the supplied forward coefficients with $F_A$ and
$F_S$ and checks their ratio in units of the respective tree
amplitudes.
The reproduction of the earlier $gg\to AA$ result and the
double-pole checks are read from the supplied run log, not
recomputed by this script.
The code used to generate the one-loop $gg\to HH$ coefficients is not included and is available from the author on request.

%% file: sections/app_constants.tex
\section{Endpoint constants}
\label{app:constants}

The boundary values of the expansion contain the constants
\begin{equation}
 \mathrm{MB}[p,q,a,w]
 =\frac{a}{(p-1)!}\int_0^1\dd t\,
 \frac{[-\ln t]^{p-1}}{1-at}\,\Li_q(wt),
 \label{eq:mb-definition}
\end{equation}
with $\Li_1(z)=-\ln(1-z)$.
Here $p$ and $q$ are positive integers, $p-1$ being the power of the logarithm and $q$ the weight of the polylogarithm, so $\mathrm{MB}[p,q,a,w]$ has weight $p+q$; the constants that occur have $p,q\in\{1,2\}$, and $a$ and $w$ take the values $-1$, $\tfrac12$ and $1$.
Besides the elementary values $\mathrm{MB}[2,2,1,1]=\pi^4/120$ and $\mathrm{MB}[2,1,1,1]=\zeta_3$, \ConstantCountWord{} distinct constants occur.
They were evaluated by quadrature at high precision, independently of the amplitude data, and reduced to the basis of Eq.~\eqref{eq:constant-basis} with the PSLQ algorithm~\cite{Ferguson:1999} at \PSLQDigits{} digits. Table~\ref{tab:mb-constants} lists the results; the largest residual is \PSLQResidual; the same relations were obtained in Mathematica.
\input{tables/mb_constants.tex}

%% file: tables/mb_constants.tex
\begin{table}[t]
\centering

\small
\begin{tabular}{cl}
\toprule
$[p,q,a,w]$ & reduction \\
\midrule
$[1,2,-1,1]$ & $- \frac{\pi^{2} \ln 2}{6} + \frac{5 \zeta_3}{8}$ \\
$[1,2,1/2,1/2]$ & $- \frac{\zeta_3}{4} - \frac{\ln^{3}2}{6} + \frac{\pi^{2} \ln 2}{12}$ \\
$[1,2,1/2,1]$ & $- \zeta_3 + \frac{\pi^{2} \ln 2}{4}$ \\
$[2,1,-1,-1]$ & $\frac{\zeta_3}{8}$ \\
$[2,1,-1,1]$ & $- \frac{13 \zeta_3}{8} + \frac{\pi^{2} \ln 2}{4}$ \\
$[2,1,1,-1]$ & $- \frac{\pi^{2} \ln 2}{4} + \zeta_3$ \\
$[2,1,1,1/2]$ & $- \frac{5 \zeta_3}{8} + \frac{\pi^{2} \ln 2}{6}$ \\
$[2,1,1/2,1/2]$ & $- \frac{\ln^{3}2}{6} + \frac{\zeta_3}{8}$ \\
$[2,1,1/2,1]$ & $- \frac{\pi^{2} \ln 2}{4} + \frac{13 \zeta_3}{8}$ \\
$[2,2,1,-1]$ & $- \frac{71 \pi^{4}}{1440} - \frac{\pi^{2} \ln^{2}2}{6} + \frac{\ln^{4}2}{6} + 4 \Li_4(\tfrac12) + \frac{7 \ln 2 \zeta_3}{2}$ \\
$[2,2,1,1/2]$ & $- \Li_4(\tfrac12) - \frac{\ln 2 \zeta_3}{4} - \frac{\pi^{2} \ln^{2}2}{24} - \frac{\ln^{4}2}{24} + \frac{19 \pi^{4}}{1440}$ \\
$[2,2,1/2,1/2]$ & $- \frac{\pi^{2} \ln^{2}2}{24} + \frac{\ln^{4}2}{24} + \frac{\pi^{4}}{1440} + \frac{\ln 2 \zeta_3}{4}$ \\
$[2,2,1/2,1]$ & $- 3 \Li_4(\tfrac12) - \frac{\pi^{2} \ln^{2}2}{8} - \frac{\ln^{4}2}{8} + \frac{7 \pi^{4}}{288}$ \\
\bottomrule
\end{tabular}
\caption{Reductions of the endpoint constants of Eq.~\eqref{eq:mb-definition}, found with PSLQ at \PSLQDigits{} digits; the largest residual is \PSLQResidual.}
\label{tab:mb-constants}
\end{table}

%% file: sections/app_normalization.tex
\section{Normalisation of the released remainders}
\label{app:normalization}

Six checks establish the normalisation in Eq.~\eqref{eq:normalisation},
using the tree amplitudes, finite remainders and Mathematica code of
the original calculation~\cite{Bhattacharya:2019bmd}.

\emph{Tree level.}
The projections of the two tree diagrams reproduce Eq.~\eqref{eq:born}
after removing a colour factor $\Nc^2-1$ present only in the $T_1$ file.
Evaluating the diagrams directly with the vertex
$iC_G\varepsilon_{\mu\alpha\rho\beta}k_1^\mu k_2^\rho\delta^{ab}$,
obtained from Eq.~\eqref{eq:EL}, gives
$|\mathcal A_{\pm\pm}|=C_G^2|\mathcal M_1^{(0)}|$ and
$|\mathcal A_{\pm\mp}|=C_G^2|\mathcal M_2^{(0)}|$
to \FeynmanDigits{} digits.

\emph{Tree and one loop.}
At threshold, the ratio of the $\nf$ coefficient in
$\mathcal M_i^{(1)}$ to $\mathcal M_i^{(0)}$ is the same for $i=1,2$.
Its imaginary part equals $-2\pi/3$ to \CataniImagDigits{} digits
at every tested point.
The one-loop diagrams proportional to $\nf$ have no cut in the
$\hat s$ channel. This imaginary part therefore comes from the
$\nf$ part of $\beta_0\ln(-\hat s/\muR^2)$ in the subtraction term
$2I^{(1)}\mathcal M^{(0)}$ of Eq.~\eqref{eq:catani}.
The check fixes the tree normalisation and phase relative to the
one-loop remainders.

\emph{Scale dependence at one loop.}
At fixed $\bA$ and $c$, dimensional analysis requires
$\mathcal M_i^{(1)}/\mA^2$ to depend on $\mA$ and $\muR$ only
through their ratio.
The released remainders are instead independent of $\muR$ because
their eight $\ln\muR^2$ terms cancel.
Under $(\mA,\muR)\to(\lambda\mA,\lambda\muR)$,
$\mathcal M_{i,\mathrm{publ}}^{(1)}/\mA^2$ changes by
$-2\beta_0\ln\lambda^2\,\mathcal M_i^{(0)}/\mA^2$,
to a relative accuracy of \ScalingRawDev{}.
The original calculation renormalises the amplitude with $Z_{GG}^2$,
where $Z_{GG}=Z_{a_s}$ and $C_G\propto\as$.
Together with the scale factor
$(-\hat s/\muR^2)^{\epsilon/2}$ in the one-loop subtraction operator
of Eq.~\eqref{eq:catani}, renormalisation group invariance of
Eq.~\eqref{eq:perturbative-expansion} requires
\begin{equation}
 \frac{\partial\mathcal M^{(1)}}{\partial\ln\muR^2}
 =2\beta_0\,\mathcal M^{(0)} .
 \label{eq:rg-one}
\end{equation}
The first line of Eq.~\eqref{eq:normalisation} restores this dependence.
The corrected ratios $\mathcal M_i^{(1)}/\mA^2$ are invariant under
the common rescaling to a relative accuracy of \ScalingFixedDev{}.

\emph{Scale dependence at two loops.}
The subtraction operator $I^{(2)}$ implemented in the original code
differs from Eqs.~(3.3) and (3.5) of Ref.~\cite{Bhattacharya:2019bmd}
by omitting the factor $(-\hat s/\muR^2)^\epsilon$ multiplying
\begin{align}
 H_g^{(2)}
 &=\tfrac12\Bigl[
 C_A^2\bigl(-\tfrac{5}{12}-\tfrac{11}{24}\zeta_2-\tfrac12\zeta_3\bigr)
 +C_A\nf\bigl(\tfrac{29}{27}+\tfrac{\zeta_2}{12}\bigr)
 -\tfrac12C_F\nf-\tfrac{5}{27}\nf^2\Bigr].
 \label{eq:H2}
\end{align}
With this convention, the scale dependence is
\begin{equation}
 \frac{\partial\Mfull{}}{\partial\ln\muR^2}
 =3\beta_0\,\mathcal M^{(1)}
 +\bigl(2\beta_1-8H_g^{(2)}-12C_F\nf\bigr)\mathcal M^{(0)},
 \label{eq:rg-two}
\end{equation}
where
$\beta_1=\tfrac{34}{3}C_A^2-2C_F\nf-\tfrac{10}{3}C_A\nf$.
The term $-12C_F\nf\mathcal M^{(0)}$ comes from the counterterm
of the $O_G$--$O_J$ amplitude, Eq.~\eqref{eq:gj-proportional},
in the renormalisation of Ref.~\cite{Bhattacharya:2019bmd}.

We fitted
$\partial\Mfull{}/\partial\ln\muR^2
=a\,\beta_0\mathcal M^{(1)}+b\,\mathcal M^{(0)}$
to the series at two kinematic points, for
$(\Nc,\nf,\mA)=$ \RGSets{}.
With both conversions in Eq.~\eqref{eq:normalisation}, the fit gives
$a=\RGCoefficient$ to \RGaDev{} and agreement for $b$ with
Eq.~\eqref{eq:rg-two} to \RGbDev{}, the accuracy of the series
at these velocities.
The second derivative agrees with $6\beta_0^2\mathcal M^{(0)}$,
as implied by Eqs.~\eqref{eq:rg-one} and \eqref{eq:rg-two},
to \RGSecondDev{}.
Without division by \FileFactor{}, the fit gives
$a=\RGCoefficientPublished$.
Without the one-loop scale term, $b$ differs from
Eq.~\eqref{eq:rg-two} by
$6\beta_0^2\ln(\mA^2/\mathrm{GeV}^2)$.
The agreement also fixes the reference scale of the released
one-loop remainders at $\muR=1$~GeV.

\emph{Origin of the factor \FileFactor{}.}
The original code multiplies the $O_G$--$O_G$ amplitude by
$\FileFactor=\FileFactorRoot^2$ during two-loop renormalisation,
including the tree and one-loop amplitudes used in IR  subtraction.
The one-loop step does not apply this factor.
The pole cancellation of Ref.~\cite{Bhattacharya:2019bmd} is unaffected
because all terms of the two-loop subtraction carry the same factor.
The accompanying code and data use the normalisation of
Eq.~\eqref{eq:normalisation}.

\emph{The original output.}
The Mathematica output of the original calculation confirms both
conversions.
Its one-loop remainders, normalised to the tree amplitude, have the
scale dependence of Eq.~\eqref{eq:rg-one} and agree with the first
line of Eq.~\eqref{eq:normalisation} to \ArchiveOneLoopDev{}.
The released one-loop files coincide with these outputs at
$\muR=1$~GeV.

An earlier two-loop renormalisation run, without the factor
\FileFactor{} and with the $O_J$ contribution switched off,
was evaluated at $v=1/11$, $z=3/11$.
Multiplying this result by \FileFactor{} and the tree amplitude,
then adding the $O_G$--$O_J$ counterterm, reproduces the real part
of the released $\mathcal M_{1,\mathrm{publ}}^{(2)}$ remainder to \ArchiveTwoLoopDev{}.
The imaginary parts differ by $32C_F$ times the one-loop
$O_G$--$O_J$ amplitude, which is proportional to the tree amplitude. This run is not an alternative result.
Its renormalisation constant exceeds $Z_{GG}^2=Z_{a_s}^2$ by a
$2\beta_1\as^2/\epsilon$ pole, and the $O_G$--$O_J$ counterterm
is absent.
Its single poles therefore do not cancel with the IR subtraction
of the released version.
In the released version, cancellation of the $C_F\nf$ single pole
fixes the normalisation of this counterterm.
A June 2019 record of the original calculation requires the same
factor in both projections at \RenormPointsWord{} kinematic points.
The files and a script reproducing these comparisons are provided
with this article.

%% file: sections/app_flavour.tex
\section{Flavour decomposition}
\label{app:flavour}

The decomposition of Eq.~\eqref{eq:nf-decomposition} separates the part of
each remainder that is independent of $\nf$ from the parts proportional to
$\nf$ and $\nf^2$. To see whether these parts add or cancel in the complete
remainder, we evaluate it at $\nf=\BenchNf$:
\begin{align}
 \Mfull{i}(\BenchNf)
 &=\Mflav{i}{0}
   +\BenchNf\,\Mflav{i}{1}
   +\BenchNf^2\,\Mflav{i}{2}
 \nonumber\\
 &\equiv a_0+a_1+a_2,
 \qquad i=1,2.
 \label{eq:nf5-sum}
\end{align}
The index $i$ is suppressed in $a_r$. The factors $\BenchNf$ and
$\BenchNf^2$ are included in $a_1$ and $a_2$, so these are the three terms
of the complete remainder rather than the flavour coefficients themselves.

\begin{figure}[!ht]
 \centering
 \includegraphics[width=\columnwidth]{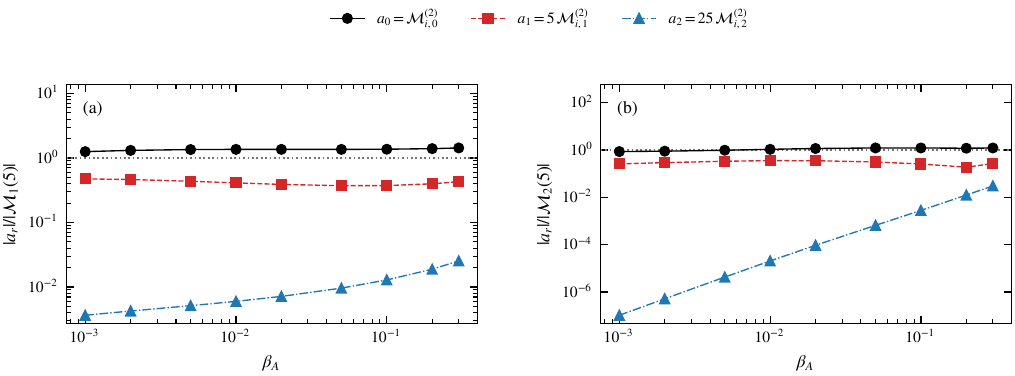}
 \caption{Magnitudes of the three terms of Eq.~\eqref{eq:nf5-sum}, divided by the magnitude of the complete coefficient: (a) $\mathcal M_1$, (b) $\mathcal M_2$, for $c=\BenchCos$, $\mA=\muR=\BenchMA\,\mathrm{GeV}$ and $\Nc=\BenchNc$. The horizontal line marks unity.}
 \label{fig:flavour-components}
\end{figure}

Figure~\ref{fig:flavour-components} shows $|a_r|/|\Mfull{i}(\BenchNf)|$ at
$c=\BenchCos$, $\Nc=\BenchNc$ and $\mA=\muR=\BenchMA\,\mathrm{GeV}$.
Since the terms are complex, the ratios need not add to unity;
a value above unity means that a term is larger in magnitude than the full remainder and is partly cancelled by the others.
The $\nf$ independent term is the largest in both coefficients over the plotted range.
The term linear in $\nf$ is not small and, for $\mathcal M_1$, partly
cancels the $\nf$-independent term. The cancellation can be quantified by
\[
 1-\frac{|a_0+a_1+a_2|}
          {|a_0|+|a_1|+|a_2|},
\]
which is zero for terms of equal phase and one for complete cancellation.
For $\mathcal M_1$ it lies between \PoneCancellationMin{} and
\PoneCancellationMax{} over the plotted range: the complete remainder is
roughly half the sum of the magnitudes of its terms. This refers to the
remainder, not the cross section. The flavour terms are therefore not a
small correction, and a remainder evaluated at one value of $\nf$ cannot
be rescaled to another. The relative sign is visible in the leading power
coefficients: $\widehat A_{12}$ depends on $\nf$ only through
$\beta_0=(11\Nc-2\nf)/3$, and $\widehat A_{11}$ and $\widehat A_{21}$
contain $\Nc a_{\mathrm A}-\nf a_{\mathrm F}$ and $4\Nc-\nf$
[Eqs.~\eqref{eq:leading-coefficients} and \eqref{eq:M1-L1}], so the quark terms
enter with the opposite sign to the gluon terms, as in the running of the coupling.  How much they cancel in the complete remainder also depends on their phases and on the powers of $\mathcal L$ they multiply. The constants $\widehat A_{10}$ and $\widehat A_{20}$ are given only numerically (Table~\ref{tab:leading-log-benchmark}), and no sign pattern is stated for them or for the power corrections; their flavour parts are included in $a_r$.

Finally, the $\nf^2$ term of $\mathcal M_2$ starts at $\bA^2(1-c^2)$ (Eq.~\eqref{eq:nf2-leading-two}).
At leading power, this term is proportional to the tree amplitude, with the same factor in both tensor structures (Sec.~\ref{sec:cancellations}), so in $\mathcal M_2$,  it shares
the centrifugal suppression of the $|J_z|=2$ tree amplitude.
This is why its relative contribution in panel~(b) becomes small near threshold.

%% file: sections/app_data.tex
\section{Evaluation time and released data}
\label{app:data}
\FloatBarrier

The series through $\bA^6$ occupy \SeriesKB~kB, against \PublishedTwoLoopMB~MB for the published two-loop files, a factor of \SizeRatio. The evaluation time was measured before the reduction of Sec.~\ref{sec:cancellations}, with the unexpanded expressions and the series in the forms used for the comparison of Sec.~\ref{sec:validation}, at $\bA=$ \TimingBetas, $c=\TimingCos$, $\mA=\muR=\TimingMA\,\mathrm{GeV}$ and $\Nc=\TimingNc$, with \TimingPrecision-digit working precision. On a first evaluation the series is \TimingFirstMin--\TimingFirstMax{} times faster. For repeated evaluations at the same point it is \TimingWarmMin--\TimingWarmMax{} times faster, with median \TimingWarmMedian; this does not correspond to a scan over new points. The two timed representations occupy \TimingSeriesMiB~MiB and \TimingExactMiB~MiB on disk, a factor of \TimingRepresentationRatio. Table~\ref{tab:timing} gives the individual measurements.
\input{tables/timing.tex}
\input{tables/computing_environment.tex}

The construction files are organised by flavour coefficient and power of $\bA$, from $\bA^{-4}$ through $\bA^6$. The series are supplied in Wolfram Language form, with numerical evaluation code in Python and Wolfram Language.
The package contains \ValidationRows{} validation records and the forward angle data.
The construction and validation code is written in Wolfram Language, C++ and Python.
The construction files and validation records are checked against the outputs of the original runs, and a script repeats the construction from the published expressions.
Each table, figure and number in this paper is generated by a single command, from these data or, for Fig.~\ref{fig:diagrams}, from its diagram source. This includes the check of the helicity sums in Fig.~\ref{fig:dps}, which starts from the supplied $gg\to HH$ coefficients, the logarithms of Sec.~\ref{sec:rate} and the normalisation checks of App.~\ref{app:normalization}.
A test suite checks the numerical evaluation code against the validation records.
Only the construction of the series from the unexpanded expressions requires Mathematica.
The series are intended for $\bA\leq\RecommendedBeta$ at the angles tested (Sec.~\ref{sec:use}).
Outside this range, use the unexpanded remainders.

%% file: tables/timing.tex
\begin{table}[t]
\centering

\footnotesize
\setlength{\tabcolsep}{2.6pt}
\begin{tabular}{lcccccc}
\toprule
& \multicolumn{3}{c}{first call [s]} & \multicolumn{3}{c}{repeated [s]} \\
$\bA$ & exact & series & ratio & exact & series & ratio \\
\midrule
$1/10$ & 56.04 & 11.61 & 4.83 & 0.3422 & 0.0178 & 19.18 \\
$1/5$ & 45.68 & 11.64 & 3.92 & 0.3403 & 0.0179 & 18.99 \\
$3/10$ & 44.64 & 11.64 & 3.84 & 0.3485 & 0.0181 & 19.22 \\
\bottomrule
\end{tabular}
\caption{Evaluation time for the two complete Type II coefficients at the benchmark parameters, $c=\TimingCos$ and \TimingPrecision-digit working precision, for the unexpanded remainders (exact) and the series. First calls start with empty caches; repeated times are medians of \TimingRepeats{} evaluations at the same point. Starting the kernel and loading the files are excluded.}
\label{tab:timing}
\end{table}

%% file: tables/computing_environment.tex
The measurements used Wolfram Language 13.2.0 for Linux x86 under Ubuntu 24.04.2 LTS on an AMD Ryzen 9 7950X processor with 124.9~GiB of memory.

%% file: main.bbl
\providecommand{\href}[2]{#2}\begingroup\raggedright\begin{thebibliography}{10}

\bibitem{Baglio:2023euv}
J.~Baglio, F.~Campanario, S.~Glaus, M.~M{\"u}hlleitner, J.~Ronca and M.~Spira,
  \emph{Full {NLO} {QCD} predictions for {Higgs}-pair production in the
  2-{Higgs}-doublet model},
  \href{https://doi.org/10.1140/epjc/s10052-023-11957-2}{\emph{Eur. Phys. J. C}
  {\bfseries 83} (2023) 826}
  [\href{https://arxiv.org/abs/2303.05409}{{\ttfamily 2303.05409}}].

\bibitem{BDFS2012}
B.~Blok, {\relax{Yu}}.~Dokshitzer, L.~Frankfurt and M.~Strikman, \emph{{pQCD}
  physics of multiparton interactions},
  \href{https://doi.org/10.1140/epjc/s10052-012-1963-8}{\emph{Eur. Phys. J. C}
  {\bfseries 72} (2012) 1963}
  [\href{https://arxiv.org/abs/1106.5533}{{\ttfamily 1106.5533}}].

\bibitem{GauntStirling2011}
J.R.~Gaunt and W.J.~Stirling, \emph{Double parton scattering singularity in
  one-loop integrals},
  \href{https://doi.org/10.1007/JHEP06(2011)048}{\emph{JHEP} {\bfseries 06}
  (2011) 048} [\href{https://arxiv.org/abs/1103.1888}{{\ttfamily 1103.1888}}].

\bibitem{DixonEsterlis2016}
L.J.~Dixon and I.~Esterlis, \emph{All orders results for self-crossing {Wilson}
  loops mimicking double parton scattering},
  \href{https://doi.org/10.1007/JHEP07(2016)116}{\emph{JHEP} {\bfseries 07}
  (2016) 116} [\href{https://arxiv.org/abs/1602.02107}{{\ttfamily
  1602.02107}}].

\bibitem{DixonEsterlis2016Erratum}
L.J.~Dixon and I.~Esterlis, \emph{Erratum: All orders results for self-crossing
  {Wilson} loops mimicking double parton scattering},
  \href{https://doi.org/10.1007/JHEP08(2016)131}{\emph{JHEP} {\bfseries 08}
  (2016) 131}.

\bibitem{Plehn:1996wb}
T.~Plehn, M.~Spira and P.M.~Zerwas, \emph{Pair production of neutral {Higgs}
  particles in gluon-gluon collisions},
  \href{https://doi.org/10.1016/0550-3213(96)00418-X}{\emph{Nucl. Phys. B}
  {\bfseries 479} (1996) 46}
  [\href{https://arxiv.org/abs/hep-ph/9603205}{{\ttfamily hep-ph/9603205}}].

\bibitem{Plehn:1996wbErratum}
T.~Plehn, M.~Spira and P.M.~Zerwas, \emph{Erratum: Pair production of neutral
  {Higgs} particles in gluon-gluon collisions}, {\emph{Nucl. Phys. B}
  {\bfseries 531} (1998) 655}.

\bibitem{DawsonDittmaierSpira1998}
S.~Dawson, S.~Dittmaier and M.~Spira, \emph{Neutral {Higgs}-boson pair
  production at hadron colliders: {QCD} corrections},
  \href{https://doi.org/10.1103/PhysRevD.58.115012}{\emph{Phys. Rev. D}
  {\bfseries 58} (1998) 115012}
  [\href{https://arxiv.org/abs/hep-ph/9805244}{{\ttfamily hep-ph/9805244}}].

\bibitem{Hespel:2014sla}
B.~Hespel, D.~L{\'o}pez-Val and E.~Vryonidou, \emph{{Higgs} pair production via
  gluon fusion in the two-{Higgs}-doublet model},
  \href{https://doi.org/10.1007/JHEP09(2014)124}{\emph{JHEP} {\bfseries 09}
  (2014) 124} [\href{https://arxiv.org/abs/1407.0281}{{\ttfamily 1407.0281}}].

\bibitem{Grober:2017gut}
R.~Gr{\"o}ber, M.~M{\"u}hlleitner and M.~Spira, \emph{{Higgs Pair Production at
  NLO QCD for CP-violating Higgs Sectors}},
  \href{https://doi.org/10.1016/j.nuclphysb.2017.10.002}{\emph{Nucl. Phys. B}
  {\bfseries 925} (2017) 1} [\href{https://arxiv.org/abs/1705.05314}{{\ttfamily
  1705.05314}}].

\bibitem{Catani:1998bh}
S.~Catani, \emph{The singular behaviour of {QCD} amplitudes at two-loop order},
  \href{https://doi.org/10.1016/S0370-2693(98)00332-3}{\emph{Phys. Lett. B}
  {\bfseries 427} (1998) 161}
  [\href{https://arxiv.org/abs/hep-ph/9802439}{{\ttfamily hep-ph/9802439}}].

\bibitem{Bhattacharya:2019bmd}
A.~Bhattacharya, M.~Mahakhud, P.~Mathews and V.~Ravindran, \emph{Two loop {QCD}
  amplitudes for di-pseudo scalar production in gluon fusion},
  \href{https://doi.org/10.1007/JHEP02(2020)121}{\emph{JHEP} {\bfseries 02}
  (2020) 121} [\href{https://arxiv.org/abs/1909.08993}{{\ttfamily
  1909.08993}}].

\bibitem{BorowkaEtAl2016}
S.~Borowka, N.~Greiner, G.~Heinrich, S.P.~Jones, M.~Kerner, J.~Schlenk et~al.,
  \emph{{Higgs} boson pair production in gluon fusion at {NLO} with full
  top-quark mass dependence},
  \href{https://doi.org/10.1103/PhysRevLett.117.012001}{\emph{Phys. Rev. Lett.}
  {\bfseries 117} (2016) 012001}
  [\href{https://arxiv.org/abs/1604.06447}{{\ttfamily 1604.06447}}].

\bibitem{BorowkaEtAl2016Erratum}
S.~Borowka, N.~Greiner, G.~Heinrich, S.P.~Jones, M.~Kerner, J.~Schlenk et~al.,
  \emph{Erratum: {Higgs} boson pair production in gluon fusion at
  next-to-leading order with full top-quark mass dependence [{Phys. Rev. Lett.}
  117, 012001 (2016)]},
  \href{https://doi.org/10.1103/PhysRevLett.117.079901}{\emph{Phys. Rev. Lett.}
  {\bfseries 117} (2016) 079901}.

\bibitem{Baglio:2020ini}
J.~Baglio, F.~Campanario, S.~Glaus, M.~M{\"u}hlleitner, J.~Ronca, M.~Spira
  et~al., \emph{{Higgs-Pair Production via Gluon Fusion at Hadron Colliders:
  NLO QCD Corrections}},
  \href{https://doi.org/10.1007/JHEP04(2020)181}{\emph{JHEP} {\bfseries 04}
  (2020) 181} [\href{https://arxiv.org/abs/2003.03227}{{\ttfamily
  2003.03227}}].

\bibitem{deFlorianMazzitelli2013NNLO}
D.~de~Florian and J.~Mazzitelli, \emph{{Higgs} boson pair production at
  next-to-next-to-leading order in {QCD}},
  \href{https://doi.org/10.1103/PhysRevLett.111.201801}{\emph{Phys. Rev. Lett.}
  {\bfseries 111} (2013) 201801}
  [\href{https://arxiv.org/abs/1309.6594}{{\ttfamily 1309.6594}}].

\bibitem{Banerjee:2018lfq}
P.~Banerjee, S.~Borowka, P.K.~Dhani, T.~Gehrmann and V.~Ravindran,
  \emph{Two-loop massless {QCD} corrections to the {$g+g\to H+H$} four-point
  amplitude}, \href{https://doi.org/10.1007/JHEP11(2018)130}{\emph{JHEP}
  {\bfseries 11} (2018) 130}
  [\href{https://arxiv.org/abs/1809.05388}{{\ttfamily 1809.05388}}].

\bibitem{ChenEtAl2020}
L.-B.~Chen, H.T.~Li, H.-S.~Shao and J.~Wang, \emph{{Higgs} boson pair
  production via gluon fusion at {N$^3$LO} in {QCD}},
  \href{https://doi.org/10.1016/j.physletb.2020.135292}{\emph{Phys. Lett. B}
  {\bfseries 803} (2020) 135292}
  [\href{https://arxiv.org/abs/1909.06808}{{\ttfamily 1909.06808}}].

\bibitem{Bhattacharya:2025hh}
A.~Bhattacharya, F.~Campanario, S.~Carlotti, J.~Chang, J.~Mazzitelli,
  M.~M{\"u}hlleitner et~al., \emph{{Higgs}-pair production via gluon fusion:
  top-yukawa- and light-quark-induced electroweak corrections},
  \href{https://doi.org/10.1007/JHEP07(2026)057}{\emph{JHEP} {\bfseries 07}
  (2026) 057} [\href{https://arxiv.org/abs/2512.14823}{{\ttfamily
  2512.14823}}].

\bibitem{H:2026ksi}
A.A.~H et~al., \emph{{Higgs Boson Pair Production via Gluon Fusion:
  Higher-Order Corrections and Theoretical Uncertainties}},
  \href{https://arxiv.org/abs/2609.04868}{{\ttfamily 2609.04868}}.

\bibitem{Brivio:2026alp}
I.~Brivio, S.~Meoni and D.~Pagani, \emph{{ALP pair production at the LHC}},
  \href{https://arxiv.org/abs/2607.21712}{{\ttfamily 2607.21712}}.

\bibitem{Chetyrkin:1998mw}
K.G.~Chetyrkin, B.A.~Kniehl, M.~Steinhauser and W.A.~Bardeen, \emph{{Effective
  QCD interactions of CP odd Higgs bosons at three loops}},
  \href{https://doi.org/10.1016/S0550-3213(98)00594-X}{\emph{Nucl. Phys. B}
  {\bfseries 535} (1998) 3}
  [\href{https://arxiv.org/abs/hep-ph/9807241}{{\ttfamily hep-ph/9807241}}].

\bibitem{Adler:1969gk}
S.L.~Adler, \emph{Axial-vector vertex in spinor electrodynamics},
  \href{https://doi.org/10.1103/PhysRev.177.2426}{\emph{Phys. Rev.} {\bfseries
  177} (1969) 2426}.

\bibitem{AdlerBardeen1969}
S.L.~Adler and W.A.~Bardeen, \emph{{Absence of higher order corrections in the
  anomalous axial vector divergence equation}},
  \href{https://doi.org/10.1103/PhysRev.182.1517}{\emph{Phys. Rev.} {\bfseries
  182} (1969) 1517}.

\bibitem{Larin1993}
S.A.~Larin, \emph{The renormalization of the axial anomaly in dimensional
  regularization},
  \href{https://doi.org/10.1016/0370-2693(93)90053-K}{\emph{Phys. Lett. B}
  {\bfseries 303} (1993) 113}
  [\href{https://arxiv.org/abs/hep-ph/9302240}{{\ttfamily hep-ph/9302240}}].

\bibitem{HarlanderKilgore2002}
R.V.~Harlander and W.B.~Kilgore, \emph{Production of a pseudo-scalar {Higgs}
  boson at hadron colliders at next-to-next-to-leading order},
  \href{https://doi.org/10.1088/1126-6708/2002/10/017}{\emph{JHEP} {\bfseries
  10} (2002) 017} [\href{https://arxiv.org/abs/hep-ph/0208096}{{\ttfamily
  hep-ph/0208096}}].

\bibitem{Sterman:1986aj}
G.~Sterman, \emph{Summation of large corrections to short-distance hadronic
  cross sections},
  \href{https://doi.org/10.1016/0550-3213(87)90258-6}{\emph{Nucl. Phys. B}
  {\bfseries 281} (1987) 310}.

\bibitem{Catani:1989ne}
S.~Catani and L.~Trentadue, \emph{Resummation of the {QCD} perturbative series
  for hard processes},
  \href{https://doi.org/10.1016/0550-3213(89)90273-3}{\emph{Nucl. Phys. B}
  {\bfseries 327} (1989) 323}.

\bibitem{deFlorian:2012za}
D.~de~Florian and J.~Mazzitelli, \emph{A next-to-next-to-leading order
  calculation of soft-virtual cross sections},
  \href{https://doi.org/10.1007/JHEP12(2012)088}{\emph{JHEP} {\bfseries 12}
  (2012) 088} [\href{https://arxiv.org/abs/1209.0673}{{\ttfamily 1209.0673}}].

\bibitem{deFlorian:2015moa}
D.~de~Florian and J.~Mazzitelli, \emph{{Higgs} pair production at
  next-to-next-to-leading logarithmic accuracy at the {LHC}},
  \href{https://doi.org/10.1007/JHEP09(2015)053}{\emph{JHEP} {\bfseries 09}
  (2015) 053} [\href{https://arxiv.org/abs/1505.07122}{{\ttfamily
  1505.07122}}].

\bibitem{Bhattacharya:2021nsv}
A.~Bhattacharya, M.C.~Kumar, P.~Mathews and V.~Ravindran,
  \emph{Next-to-soft-virtual resummed prediction for pseudoscalar {Higgs} boson
  production at {NNLO}+{$\overline{\text{NNLL}}$}},
  \href{https://doi.org/10.1103/PhysRevD.105.116015}{\emph{Phys. Rev. D}
  {\bfseries 105} (2022) 116015}
  [\href{https://arxiv.org/abs/2112.02341}{{\ttfamily 2112.02341}}].

\bibitem{Landau:1959fi}
L.D.~Landau, \emph{On analytic properties of vertex parts in quantum field
  theory}, \href{https://doi.org/10.1016/0029-5582(59)90154-3}{\emph{Nucl.
  Phys.} {\bfseries 13} (1959) 181}.

\bibitem{BDFS2011}
B.~Blok, {\relax{Yu}}.~Dokshitzer, L.~Frankfurt and M.~Strikman, \emph{The four
  jet production at {LHC} and {Tevatron} in {QCD}},
  \href{https://doi.org/10.1103/PhysRevD.83.071501}{\emph{Phys. Rev. D}
  {\bfseries 83} (2011) 071501}
  [\href{https://arxiv.org/abs/1009.2714}{{\ttfamily 1009.2714}}].

\bibitem{Diehl:2019rdh}
M.~Diehl, J.R.~Gaunt, P.~Pl{\"o}{\ss}l and A.~Sch{\"a}fer, \emph{Two-loop
  splitting in double parton distributions},
  \href{https://doi.org/10.21468/SciPostPhys.7.2.017}{\emph{SciPost Phys.}
  {\bfseries 7} (2019) 017} [\href{https://arxiv.org/abs/1902.08019}{{\ttfamily
  1902.08019}}].

\bibitem{DiehlOstermeierSchafer2012}
M.~Diehl, D.~Ostermeier and A.~Sch{\"a}fer, \emph{Elements of a theory for
  multiparton interactions in {QCD}},
  \href{https://doi.org/10.1007/JHEP03(2012)089}{\emph{JHEP} {\bfseries 03}
  (2012) 089} [\href{https://arxiv.org/abs/1111.0910}{{\ttfamily 1111.0910}}].

\bibitem{DiehlOstermeierSchafer2012Erratum}
M.~Diehl, D.~Ostermeier and A.~Sch{\"a}fer, \emph{Erratum: Elements of a theory
  for multiparton interactions in {QCD}},
  \href{https://doi.org/10.1007/JHEP03(2016)001}{\emph{JHEP} {\bfseries 03}
  (2016) 001}.

\bibitem{DiehlSchafer2011}
M.~Diehl and A.~Sch{\"a}fer, \emph{Theoretical considerations on multiparton
  interactions in {QCD}}, {\emph{Phys. Lett. B} {\bfseries 698} (2011) 389}
  [\href{https://arxiv.org/abs/1102.3081}{{\ttfamily 1102.3081}}].

\bibitem{BDFS2014}
B.~Blok, {\relax{Yu}}.~Dokshitzer, L.~Frankfurt and M.~Strikman,
  \emph{Perturbative {QCD} correlations in multi-parton collisions},
  \href{https://doi.org/10.1140/epjc/s10052-014-2926-z}{\emph{Eur. Phys. J. C}
  {\bfseries 74} (2014) 2926}
  [\href{https://arxiv.org/abs/1306.3763}{{\ttfamily 1306.3763}}].

\bibitem{Manohar:2012pe}
A.V.~Manohar and W.J.~Waalewijn, \emph{{What is double parton scattering?}},
  \href{https://doi.org/10.1016/j.physletb.2012.05.044}{\emph{Phys. Lett. B}
  {\bfseries 713} (2012) 196}
  [\href{https://arxiv.org/abs/1202.5034}{{\ttfamily 1202.5034}}].

\bibitem{Gaunt:2012dd}
J.R.~Gaunt, \emph{{Single perturbative splitting diagrams in double parton
  scattering}}, \href{https://doi.org/10.1007/JHEP01(2013)042}{\emph{JHEP}
  {\bfseries 01} (2013) 042} [\href{https://arxiv.org/abs/1207.0480}{{\ttfamily
  1207.0480}}].

\bibitem{DiehlGauntSchoenwald2017}
M.~Diehl, J.R.~Gaunt and K.~Sch{\"o}nwald, \emph{Double hard scattering without
  double counting}, \href{https://doi.org/10.1007/JHEP06(2017)083}{\emph{JHEP}
  {\bfseries 06} (2017) 083}
  [\href{https://arxiv.org/abs/1702.06486}{{\ttfamily 1702.06486}}].

\bibitem{Buffing:2017mqm}
M.G.A.~Buffing, M.~Diehl and T.~Kasemets, \emph{{Transverse momentum in double
  parton scattering: factorisation, evolution and matching}},
  \href{https://doi.org/10.1007/JHEP01(2018)044}{\emph{JHEP} {\bfseries 01}
  (2018) 044} [\href{https://arxiv.org/abs/1708.03528}{{\ttfamily
  1708.03528}}].

\bibitem{Buffing:2017mqmErratum}
M.G.A.~Buffing, M.~Diehl and T.~Kasemets, \emph{{Erratum: Transverse momentum
  in double parton scattering: factorisation, evolution and matching}},
  \href{https://doi.org/10.1007/JHEP07(2021)046}{\emph{JHEP} {\bfseries 07}
  (2021) 046}.

\bibitem{RyskinSnigirev2011}
M.G.~Ryskin and A.M.~Snigirev, \emph{A fresh look at double parton scattering},
  \href{https://doi.org/10.1103/PhysRevD.83.114047}{\emph{Phys. Rev. D}
  {\bfseries 83} (2011) 114047}
  [\href{https://arxiv.org/abs/1103.3495}{{\ttfamily 1103.3495}}].

\bibitem{GauntMaciulaSzczurek2014}
J.R.~Gaunt, R.~Maciu{\l}a and A.~Szczurek, \emph{Conventional versus
  single-ladder-splitting contributions to double parton scattering production
  of two quarkonia, two {Higgs} bosons and {$c\bar c c\bar c$}},
  \href{https://doi.org/10.1103/PhysRevD.90.054017}{\emph{Phys. Rev. D}
  {\bfseries 90} (2014) 054017}
  [\href{https://arxiv.org/abs/1407.5821}{{\ttfamily 1407.5821}}].

\bibitem{CMS:2022pio}
{CMS Collaboration}, \emph{Observation of same-sign {WW} production from double
  parton scattering in proton-proton collisions at {$\sqrt{s}=13$}~{TeV}},
  \href{https://doi.org/10.1103/PhysRevLett.131.091803}{\emph{Phys. Rev. Lett.}
  {\bfseries 131} (2023) 091803}
  [\href{https://arxiv.org/abs/2206.02681}{{\ttfamily 2206.02681}}].

\bibitem{ATLAS:2025dps}
{ATLAS Collaboration}, \emph{Observation of double parton scattering in
  same-sign {$W$} boson pair production in {$pp$} collisions at
  {$\sqrt{s}=13$}~{TeV} with the {ATLAS} detector},
  \href{https://doi.org/10.1016/j.physletb.2025.139892}{\emph{Phys. Lett. B}
  {\bfseries 870} (2025) 139892}
  [\href{https://arxiv.org/abs/2505.08313}{{\ttfamily 2505.08313}}].

\bibitem{DiehlPloessl2026}
M.~Diehl and P.~Pl{\"o}{\ss}l, \emph{A quantitative study of two-loop splitting
  in double parton distributions},
  \href{https://doi.org/10.1007/JHEP05(2026)158}{\emph{JHEP} {\bfseries 05}
  (2026) 158} [\href{https://arxiv.org/abs/2602.18233}{{\ttfamily
  2602.18233}}].

\bibitem{SupplementalMaterial}
A.~Bhattacharya, ``Supplemental material for `{T}wo-loop $gg\to {AA}$
  amplitudes near threshold and the double parton scattering singularity'.''
  \url{https://github.com/abhattacharya1992/ggAA-threshold}, 2026.
\newblock 10.5281/zenodo.23597489.

\bibitem{Glover:1987nx}
E.W.N.~Glover and J.J.~van~der Bij, \emph{{Higgs boson pair production via
  gluon fusion}},
  \href{https://doi.org/10.1016/0550-3213(88)90083-1}{\emph{Nucl. Phys. B}
  {\bfseries 309} (1988) 282}.

\bibitem{Bell:1969ts}
J.S.~Bell and R.~Jackiw, \emph{A {PCAC} puzzle: {$\pi^0\to\gamma\gamma$} in the
  {$\sigma$}-model}, \href{https://doi.org/10.1007/BF02823296}{\emph{Nuovo Cim.
  A} {\bfseries 60} (1969) 47}.

\bibitem{Bhattacharya:2026ucv}
A.~Bhattacharya, F.~Campanario, M.~Kubocz, T.~Figy, T.~Kushimo and I.~Rosario,
  \emph{{Using Higgs Boson Production in Association with Three Jets via Gluon
  Fusion at the LHC to Investigate the CP Properties of the Higgs Boson}},
  \href{https://arxiv.org/abs/2610.37627}{{\ttfamily 2610.37627}}.

\bibitem{Frellesvig:2016ske}
H.~Frellesvig, D.~Tommasini and C.~Wever, \emph{On the reduction of generalized
  polylogarithms to {$\mathrm{Li}_n$} and {$\mathrm{Li}_{2,2}$} and on the
  evaluation thereof},
  \href{https://doi.org/10.1007/JHEP03(2016)189}{\emph{JHEP} {\bfseries 03}
  (2016) 189} [\href{https://arxiv.org/abs/1601.02649}{{\ttfamily
  1601.02649}}].

\bibitem{Vollinga:2004sn}
J.~Vollinga and S.~Weinzierl, \emph{Numerical evaluation of multiple
  polylogarithms},
  \href{https://doi.org/10.1016/j.cpc.2004.12.009}{\emph{Comput. Phys. Commun.}
  {\bfseries 167} (2005) 177}
  [\href{https://arxiv.org/abs/hep-ph/0410259}{{\ttfamily hep-ph/0410259}}].

\bibitem{RemiddiVermaseren2000}
E.~Remiddi and J.A.M.~Vermaseren, \emph{Harmonic polylogarithms},
  \href{https://doi.org/10.1142/S0217751X00000367}{\emph{Int. J. Mod. Phys. A}
  {\bfseries 15} (2000) 725}
  [\href{https://arxiv.org/abs/hep-ph/9905237}{{\ttfamily hep-ph/9905237}}].

\bibitem{Henn2013}
J.M.~Henn, \emph{Multiloop integrals in dimensional regularization made
  simple}, \href{https://doi.org/10.1103/PhysRevLett.110.251601}{\emph{Phys.
  Rev. Lett.} {\bfseries 110} (2013) 251601}
  [\href{https://arxiv.org/abs/1304.1806}{{\ttfamily 1304.1806}}].

\bibitem{BenekeSmirnov1998}
M.~Beneke and V.A.~Smirnov, \emph{Asymptotic expansion of {Feynman} integrals
  near threshold},
  \href{https://doi.org/10.1016/S0550-3213(98)00138-2}{\emph{Nucl. Phys. B}
  {\bfseries 522} (1998) 321}
  [\href{https://arxiv.org/abs/hep-ph/9711391}{{\ttfamily hep-ph/9711391}}].

\bibitem{Bellafronte:2022}
L.~Bellafronte, G.~Degrassi, P.P.~Giardino, R.~Gr{\"o}ber and M.~Vitti,
  \emph{Gluon fusion production at {NLO}: Merging the transverse momentum and
  the high-energy expansions},
  \href{https://doi.org/10.1007/JHEP07(2022)069}{\emph{JHEP} {\bfseries 07}
  (2022) 069} [\href{https://arxiv.org/abs/2202.12157}{{\ttfamily
  2202.12157}}].

\bibitem{Davies:2023}
J.~Davies, G.~Mishima, K.~Sch{\"o}nwald and M.~Steinhauser, \emph{Analytic
  approximations of {$2\to2$} processes with massive internal particles},
  \href{https://doi.org/10.1007/JHEP06(2023)063}{\emph{JHEP} {\bfseries 06}
  (2023) 063} [\href{https://arxiv.org/abs/2302.01356}{{\ttfamily
  2302.01356}}].

\bibitem{Groeber:2018}
R.~Gr{\"o}ber, A.~Maier and T.~Rauh, \emph{Reconstruction of top-quark mass
  effects in {H}iggs pair production and other gluon-fusion processes},
  \href{https://doi.org/10.1007/JHEP03(2018)020}{\emph{JHEP} {\bfseries 03}
  (2018) 020} [\href{https://arxiv.org/abs/1709.07799}{{\ttfamily
  1709.07799}}].

\bibitem{Ferguson:1999}
H.R.P.~Ferguson, D.H.~Bailey and S.~Arno, \emph{Analysis of {PSLQ}, an integer
  relation finding algorithm},
  \href{https://doi.org/10.1090/S0025-5718-99-00995-3}{\emph{Math. Comp.}
  {\bfseries 68} (1999) 351}.

\bibitem{Blumlein:2009cf}
J.~Bl{\"u}mlein, D.J.~Broadhurst and J.A.M.~Vermaseren, \emph{The multiple zeta
  value data mine},
  \href{https://doi.org/10.1016/j.cpc.2009.11.007}{\emph{Comput. Phys. Commun.}
  {\bfseries 181} (2010) 582}
  [\href{https://arxiv.org/abs/0907.2557}{{\ttfamily 0907.2557}}].

\bibitem{mpmath}
{The mpmath development team}, ``mpmath: a {Python} library for
  arbitrary-precision floating-point arithmetic (version 1.3.0).''
  \url{https://mpmath.org}, 2023.

\bibitem{Mangano:1990by}
M.L.~Mangano and S.J.~Parke, \emph{Multiparton amplitudes in gauge theories},
  \href{https://doi.org/10.1016/0370-1573(91)90091-Y}{\emph{Phys. Rept.}
  {\bfseries 200} (1991) 301}
  [\href{https://arxiv.org/abs/hep-th/0509223}{{\ttfamily hep-th/0509223}}].

\bibitem{Dixon:1996wi}
L.J.~Dixon, \emph{Calculating scattering amplitudes efficiently},  in
  \emph{{QCD} and beyond: Proceedings of {TASI} 95}, pp.~539--584, 1996
  [\href{https://arxiv.org/abs/hep-ph/9601359}{{\ttfamily hep-ph/9601359}}].

\bibitem{KorchemskayaKorchemsky1995}
I.A.~Korchemskaya and G.P.~Korchemsky, \emph{High-energy scattering in {QCD}
  and cross singularities of {W}ilson loops}, {\emph{Nucl. Phys. B} {\bfseries
  437} (1995) 127} [\href{https://arxiv.org/abs/hep-ph/9409446}{{\ttfamily
  hep-ph/9409446}}].

\bibitem{deFlorianMazzitelli2013Virtual}
D.~de~Florian and J.~Mazzitelli, \emph{Two-loop virtual corrections to {Higgs}
  pair production},
  \href{https://doi.org/10.1016/j.physletb.2013.06.046}{\emph{Phys. Lett. B}
  {\bfseries 724} (2013) 306}
  [\href{https://arxiv.org/abs/1305.5206}{{\ttfamily 1305.5206}}].

\end{thebibliography}\endgroup
